%% file: ms.tex
\documentclass[twocolumn]{aastex631}
\usepackage{CJK}
\usepackage{xcolor}
\usepackage{amsmath}
\usepackage{savesym}

\usepackage[utf8]{inputenc}
\def\ra#1#2#3{#1$^{\rm h}$#2$^{\rm m}$#3$^{\rm s}$}
\def\dec#1#2#3{$#1^\circ#2'#3''$}

\shorttitle{Obscured Star Formation in the Host of FRB~20201124A}
\shortauthors{Dong et al.}

\newcommand{\kms}{km\,s$^{-1}$}

\begin{document}
\begin{CJK*}{UTF8}{gbsn}

\title{Mapping Obscured Star Formation in the Host Galaxy of FRB\,20201124A}

\correspondingauthor{Y. Dong}
\email{yuxin.dong@northwestern.edu}

\input{aff.tex}
\input{authors.tex}

\begin{abstract}
We present high-resolution 1.5--6~GHz Karl G. Jansky Very Large Array (VLA) and \textit{Hubble Space Telescope} (\textit{HST}) optical and infrared observations of the extremely active repeating fast radio burst (FRB) FRB\,20201124A and its barred spiral host galaxy. We constrain the location and morphology of star formation in the host and search for a persistent radio source (PRS) coincident with FRB\,20201124A. We resolve the morphology of the radio emission across all frequency bands and measure a star formation rate SFR $\approx 8.9\,M_{\odot}$~yr$^{-1}$, approximately $\approx 2.5-6$ times larger than optically-inferred SFRs, demonstrating dust-obscured star formation throughout the host. Compared to a sample of all known FRB hosts with radio emission, the host of FRB\,20201124A has the most significantly obscured star formation. While {\it HST} observations show the FRB to be offset from the bar or spiral arms, the radio emission extends to the FRB location. We propose that the FRB progenitor could have formed \textit{in situ} (e.g., a magnetar born from a massive star explosion). It is still plausible, although less likely, that the progenitor of FRB\,20201124A migrated from the central bar of the host. We further place a limit on the luminosity of a putative PRS at the FRB position of $L_{\rm 6.0 \ GHz}$ $\lesssim$ 1.8 $\times 10^{27}$\,erg\,s$^{-1}$\,Hz$^{-1}$, among the deepest PRS luminosity limits to date. However, this limit is still broadly consistent with both magnetar nebulae and hypernebulae models assuming a constant energy injection rate of the magnetar and an age of $\gtrsim 10^{5}$ yr in each model, respectively.
\end{abstract}

\section{Introduction} \label{sec:intro}
Fast Radio Bursts (FRBs) are dispersed, millisecond duration bright radio pulses, located primarily at cosmological distances \citep{Lorimer07, Cordes19, Petroff22, Zhang22}. Among the hundreds of FRBs that have been identified \citep{CHIMB}, the vast majority are observed as apparent one-off events (so-called ``non-repeaters", \citealt{Shannon18, CHIMB}), while only a handful exhibit repeat bursts, known as ``repeaters" \citep{Spitler16}. With only a small number of precisely localized events and an apparent dichotomy between repeating and non-repeating events, the physical origin(s) of FRBs remain an enigma. 

A number of FRB progenitor models exist, with the bulk of models connected with young and rapidly spinning neutron stars with extremely strong magnetic fields (``magnetars'', \citealt{Margalit19, Gourdji20}). This connection was strengthened by the detection of an FRB-like event (FRB\,20200428A) from the Galactic magnetar SGR\,1935+2154 \citep{Bochenek20, CHIME-FRB200428}. However, existing progenitor models are challenged by the emerging diversity of FRB host environments. While the majority of hosts are star-forming \citep{Chatterjee17, Niu22, Bhandari23, Gordon2023}, a subset are quiescent and massive galaxies \citep{Li20, Gordon2023, Sharma23}. Characterizing the immediate environments of FRBs have led to the discovery of the repeating FRB\,20200120E coincident with an old globular cluster in the nearby M81 \citep{Bhardwaj21_M81, Kirsten22}, whereas several repeating and non-repeating FRBs were found coincident with the spiral arms of their host galaxies \citep{Mannings21,Tendulkar21}. Moreover, while compact, persistent radio sources (PRSs) were discovered coincident with two repeating FRBs located in low-mass dwarf galaxies \citep{Chatterjee17,Niu22}, no additional PRS have been discovered to date despite a number of efforts \citep{Law22,Law23}. The observed heterogeneity among FRB hosts and their local environments suggests that FRB sources may be produced via multiple progenitor systems or formation channels. 

While the population of known hosts has grown to $\sim 2$ dozen events, only a few have been localized to milliarcsecond precision, paving the way for detailed studies of their parsec-scale environments. The repeating FRB\,20201124A is one such event, and was first reported by the Canadian Hydrogen Intensity Mapping Experiment FRB collaboration (CHIME/
FRB; \citealt{CHIMB}) on 2020 November 24 UTC 08:50:41. Following its initial discovery, the FRB has been observed to exhibit periods of heightened activity with hundreds of bursts recorded over several months \citep{Lanman22, Xu22}. An interferometric subarcsecond location for the source was first determined by \cite{Day21} using the Australian Square Kilometre Array Pathfinder (ASKAP) and subsequently improved to milliarcsecond precision with the European Very Large Baseline Interferometry (VLBI) Network (EVN; \citealt{Marcote21, Nimmo22}). Thus far, over 2500 distinct bursts have been reported by numerous radio facilities, including the Upgraded Giant Metrewave Radio Telescope (uGMRT; \citealt{Wharton21}), the Five-hundred-meter Aperture Spherical radio Telescope (FAST; \citealt{Xu22, Zhou22}), and the Karl G. Jansky Very Large Array (VLA; \citealt{Law2021}), establishing FRB\,20201124A as one of the most prolific repeating FRBs to date \citep{Kirsten23}.

FRB\,20201124A was pinpointed to its host galaxy at $z=0.0979$, and follow-up observational efforts have uncovered faint and extended radio emission centered on the host and interpreted as star formation \citep{Fong21, Ricci21, Ravi22, Xu22}. In contrast, deep observations at 22~GHz showed extended radio emission offset from the host center and co-located with the FRB site, suggesting that the FRB progenitor formed \textit{in-situ} in a star forming region \citep{Piro21}. This is also supported by the detection of resolved radio emission at the VLBI scale with the EVN \citep{Marcote21,Nimmo22}. However, ground-based optical observations of the host galaxy show that FRB\,20201124A is offset from any region of apparent star formation (i.e., spiral arms and bar;  \citealt{Xu22}) challenging the notion of \textit{in-situ} progenitor formation.

One way to reconcile this discrepancy is by obtaining high-resolution imaging of the FRB host to map its true morphology and search for signs of obscured star formation. Here, we present multiwavelength follow-up observations of FRB\,20201124A. We report a compilation of VLA, \textit{HST}, and Keck observations of the host galaxy of FRB\,20201124A in section \ref{sec:obs}. In Section~\ref{sec:res},  we present an analysis of the location and morphology of star formation in the host and constrain the amount of dust obscuration within the host environment. We also place constraints on the luminosity of a putative PRS coincident with FRB\,20201124A. In Section~\ref{sec:discuss}, we discuss implications for the FRB progenitor, the level of observed dust obscuration, and the nature of the putative PRS in the context of current progenitor models. We summarize our results and conclude in Section~\ref{sec:conclude}. Throughout the paper, we adopt the \textit{Planck} cosmological parameters \citep{Planck18} where $H_{0}$ = 67.66 km s$^{-1}$ Mpc$^{-1}$, $\Omega_m = 0.310$, and $\Omega_{\lambda} = 0.690$.

\input{tables/radio_obs.tex} 

\section{Observations} \label{sec:obs}

\subsection{Radio Observations}

We obtained observations of the field of FRB\,20201124A with the VLA under program 22A-213 (PI: W.~Fong) in A-configuration with a maximum baseline length of 36 km. The field was observed for a total of $\sim$ 3.5 hours between 2022 March 6 and March 7 UTC in three frequency bands centered at 1.5 (L-band), 3.0 (S-band), and 6.0~GHz (C-band). We utilized the 3-bit samplers which provide the full 4~GHz of bandwidth across the observing band at high frequencies (4-8 GHz) and the 8-bit samplers with 1 and 2~GHz bandwidth in the lower frequency bands. Due to radio frequency interference (RFI) and the excision of edge channels, the effective bandwidths are 0.8~GHz, 1.6~GHz, and 3.6~GHz, in L-, S-, and C-bands, respectively. We performed bandpass and flux density calibration using 3C147 and complex gain calibration using J0510+1800. 

The data were processed using the standard VLA calibration pipeline (version 2022.2.0.64) as part of the Common Astronomy Software Applications (CASA, \citealt{casa, CASA22})software package. During our data reduction process, we encountered two main issues: gain compression caused by strong RFI signals, and a problematic gain calibration source. The second issue was caused by the intra-day variable nature of our complex gain calibrator, which is a very compact blazar source. This leads to both a complicated frequency spectrum that is poorly described by a low-order polynomial, and potential temporal variability during our observation (although based on historic data, we do not expect the source to vary by more than 5$\%$ on timescales of a couple of hours; \citealt{Koay11}). We estimate that these issues could collectively contribute a systematic error in the flux density scale at up to a 30$\%$ level. To mitigate both issues as much as possible, we re-ran the VLA pipeline after enabling gain compression correction and revising the gain calibrator polynomial fit order to 4.

We imaged the field with CASA's \texttt{tclean} task out to the first null of the primary beam in each band. We used a pixel scale of 0.26, 0.13, and 0.066\arcsec/pixel at 1.5, 3, and 6~GHz, respectively. We performed deconvolution using widefield gridders, a natural visibility weighting scheme, and multi-term multi-frequency synthesis (MTMFS; \citealt{RC11}) with two Taylor terms. We also imaged the data using a robust weighting scheme which obtains $\sim40\%$ higher angular resolution at the cost of $\sim30\%$ higher noise; the higher angular resolution robust images are used predominantly to search for compact continuum emission at the site of the FRB (Section~\ref{sec:PRS}). We performed a wideband primary beam (PB) correction using \texttt{widebandpbcor} to account for the falling sensitivity away from the phase center. The image is divided by the PB correction map using two Taylor terms and a minimum gain level of 0.01. To measure the level of any flux offset due to the problematic gain calibrator, we determined the flux densities of nearby point sources in the field at 3~GHz and compared them with archival values from the Very Large Array Sky Survey (VLASS; \citealt{VLASS}). We find that our measured flux densities are of order $\sim$10$\%$ higher than the archival values -- the limited signal--to--noise ratio of the archival VLASS observations precluding a more precise estimate. Thus, we conclude that despite our efforts to correct for the effects of our imperfect gain calibrator source, the flux scale of our images of FRB~20201124A are likely slightly high. As a result, we treat the measured fluxes with caution when comparing them to archival observations (Section~\ref{sec:globalenv}).

We also obtained and re-imaged the VLA observations of FRB\,20201124A at 22~GHz (Program SG9112; PI: L.~Piro), originally presented in \citet{Piro21}. In their work, the 22~GHz data yield a detection that exhibits an extended morphology possibly co-located with the FRB position \citep{Piro21} and hence is useful as a comparison to our high-resolution data in this work. For these observations, we used standard gridders, $9000 \times 9000$ pixels, and a pixel scale of 0.19\arcsec/pixel while keeping all other parameters the same. To ensure there are no large astrometric offsets across observations, we checked both the calibrator position precision and a field source in the 22~GHz image. We find that the uncertainty is at most 0.18\arcsec, i.e. of the order of the pixel scale, thereby confirming that the astrometric offset is small.

\begin{figure*}
\makebox[\textwidth][c]{\includegraphics[width=\textwidth]{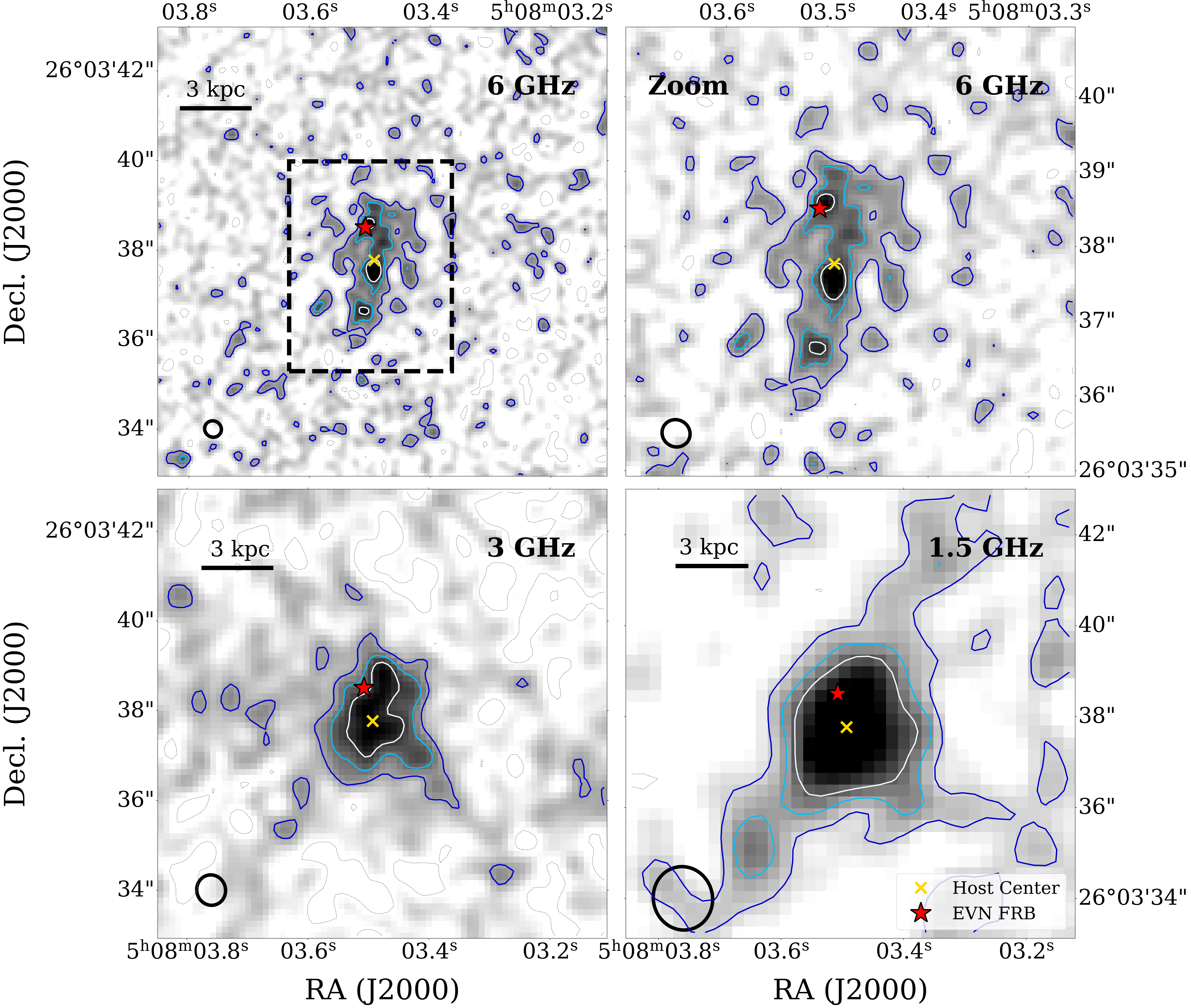}}
\caption{VLA images with natural weighting of the host galaxy of \textsc{FRB\,20201124A} taken in the extended A-array configuration in three bands: 6 GHz (top-left; 10\arcsec $\times$ 10\arcsec, top-right; zoomed in, 6\arcsec $\times$ 6\arcsec), 3~GHz (bottom-left; 10\arcsec $\times$ 10\arcsec) and 1.5~GHz (bottom-right; 10\arcsec $\times$ 10\arcsec). The beam size is displayed in the bottom left corner at each frequency. In each panel, the FRB position is represented as a red star and the host centroid determined from {\it HST} imaging is marked as a yellow cross. Contours denote -2$\sigma$ (dotted, grey), 2$\sigma$ (dark blue), 4$\sigma$ (light blue), and 6$\sigma$ (white) significance levels, where $\sigma=13.95~\mu$Jy/beam at 1.5 GHz, 6.29 $\mu$Jy/beam at 3 GHz, and 1.94 $\mu$Jy/beam at 6 GHz. The radio emission is roughly aligned with the host center in all frequencies. While we are unable to resolve the smaller-scale structures at 1.5 and 3 GHz, the 6~GHz emission clearly exhibits complex morphology, with one peak located near the FRB position and another closer to the host galaxy center.
}
\label{fig:radio}
\end{figure*}

For all three bands, we employed phase-only self-calibration to recover lost flux that is otherwise de-correlated due to atmospheric fluctuations, and to further improve the dynamic range. We average over all spectral windows, scans, and polarizations to obtain the most robust solutions at each frequency band, and use a minimum signal-to-noise ratio (SNR) threshold of 8. From the self-calibrated images, we extracted the flux density of the source using the \texttt{BLOBCAT} package \citep{BLOBCAT}. \texttt{BLOBCAT} utilizes a flood-fill algorithm to estimate the total flux density of sources of complex morphology. We set the minimum peak SNR to be 5$\sigma$ and the lower limit for ``flooding" contiguous pixels adjoining an initial peak to be 2.5$\sigma$. We detected extended radio emission coincident with the host galaxy and the position of FRB\,20201124A in all frequency bands (Figure~\ref{fig:radio}) and determined flux densities of $F_{\nu}$ = $750 \pm 114$ $\mu$Jy at 1.5~GHz, $547 \pm 82$ $\mu$Jy at 3~GHz, and $366 \pm 55$ $\mu$Jy at~6 GHz. To determine regions of source significance relative to the root mean square (RMS, $\sigma$) of the image, we defined the RMS as the noise level of the image calculated with pixel statistics in \texttt{SAOImage DS9} and determined contours corresponding to significance levels at -2, 2, 4, and 6$\sigma$. The contours in all three bands, along with the corresponding images, are plotted in Figure~\ref{fig:radio}.  The relatively large uncertainty on the total flux density at 6~GHz is a result of the highly resolved nature of the source, as much of the extended emission is detected only at low significance. The details of our radio observations and derived flux densities are summarized in Table \ref{tab:radio_obs}.

\begin{figure*}
\makebox[\textwidth][c]{\includegraphics[width=\textwidth]{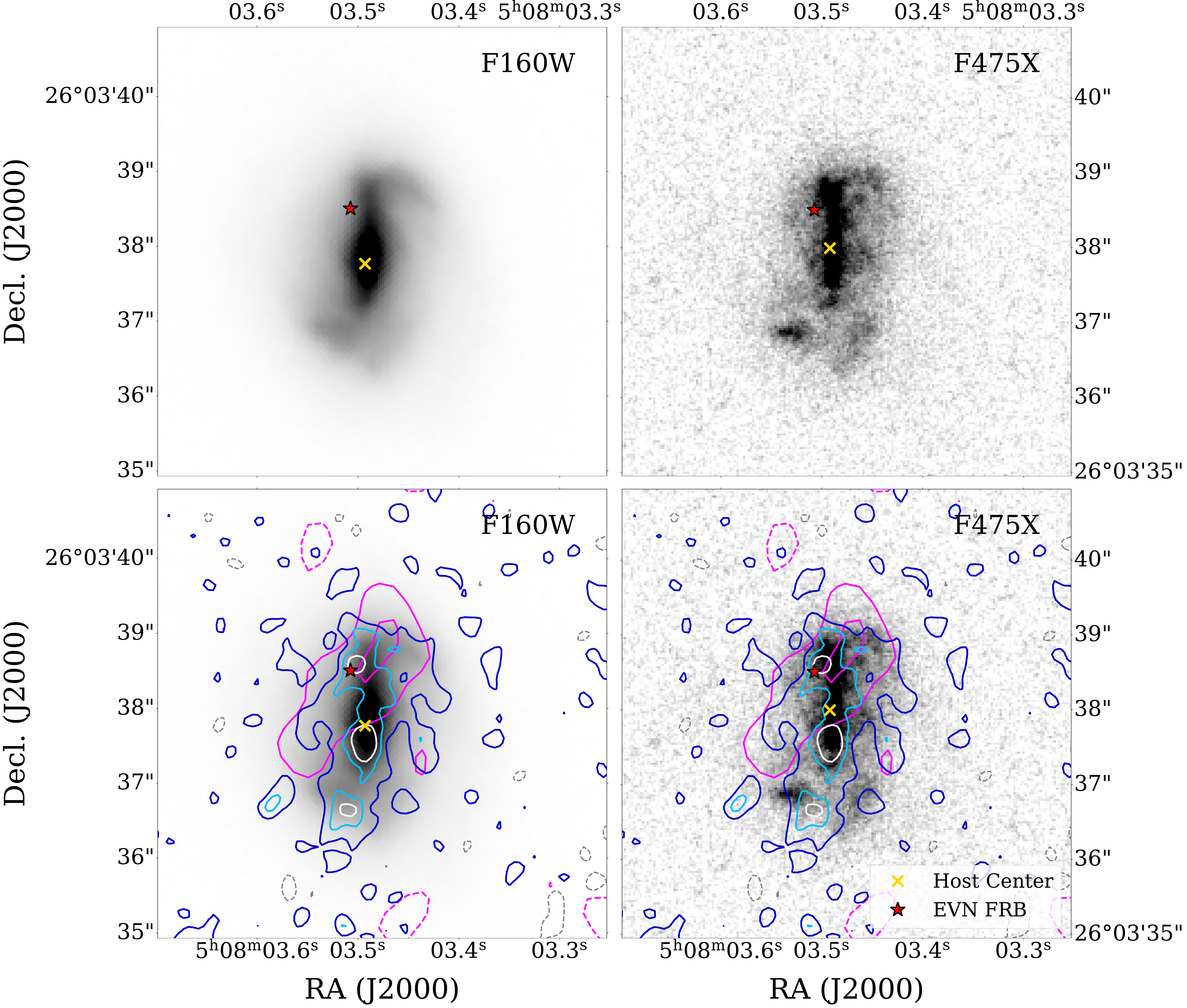}}
\caption{\textit{HST} imaging of the host galaxy of FRB\,20201124A in filters F160W (left; IR) and F475X (right; optical) showing a barred spiral morphology. All panels are 6\arcsec $\times$ 6\arcsec. The host center is denoted as a yellow cross and the FRB EVN position is marked as a red star. Radio contours at 6~GHz are overlaid at -2$\sigma$ (dotted, grey), 2$\sigma$ (dark blue), 4$\sigma$ (light blue), and 6$\sigma$ (white) significance in the bottom panels. Contours for the elongated radio emission at 22 GHz as observed by \cite{Piro21} are shown at -2, 2, and 4$\sigma$ significance in magenta.}
\label{fig:optical}
\end{figure*}

At the lower frequencies (1.5~GHz and 3~GHz) where the emission is only moderately resolved, the reduced angular resolution smooths away small-scale variations in morphology (see Figure~\ref{fig:radio}, bottom panels). At 6~GHz, the source is clearly resolved and exhibits a complex morphology, with extended emission tracing the structures seen at lower frequencies, but with (low-significance) variations as seen in the top panels of Figure~\ref{fig:radio}. The brightest peak at 6~GHz is close to the center of the host galaxy and the milliarcsecond-precision localization of the FRB \citep{Nimmo22}. However, we caution that this brightest peak is only a couple of standard deviations above the local background, as discussed further in Section~\ref{sec:PRS}. The peak is broadly consistent with the 3$\sigma$ contour emission from the 22 GHz data (see Figure \ref{fig:optical}; \citealt{Piro21}).
While the overall 6~GHz emission traces the {\it HST} IR emission, the brightest region is more concentrated at the inter-arm region. We discuss the radio morphology in the context of the optical and IR morphology in Section~\ref{sec:HST_obs}.

\subsection{HST Observations}\label{sec:HST_obs}
We obtained imaging of the host galaxy of FRB\,20201124A with the Wide Field Camera~3 (WFC3) on-board the \textit{HST} under Program 16877 (PI: A.~Mannings) on 2022 August 15-16 UTC. The host was observed in the F475X and F160W filters for one orbit each using the UVIS and IR channels, respectively. The F475X filter was chosen for its high (25\%) near-ultraviolet (UV) throughput to enable studying the spatial distribution of young stars, while the F160W is the reddest wide filter available to trace relatively older stellar populations. Similar to the method used in \citet{Mannings21}, we obtained four 597s exposures ($\sim$ 40 minutes total) in F475X in order to obtain images free of cosmic rays and image artifacts, and sub-sampled the point spread function (PSF) for better pixel sampling. We used a custom 4-point dither pattern that is five times larger than the default box pattern to dither over fixed pattern noise as described in \citet{Rafelski:2015}. To maximize UV throughput and minimize our pixel-based charge transfer efficiency (CTE) correction, we placed the target on chip 2 close to the readout. The background in these exposures was sufficiently high to not require the use of post-flash. For the F160W observations, we used SPAR25 with NSAMP of 15 for a total of six exposures of 353s, each using a 6-point wide dither pattern from \citet{Anderson:2016}. The dither offsets were multiplied by 3 for improved blob, artifact, and persistence removal. The exposures were obtained at similar orients with sufficient overlap for good alignment. 

We retrieved the data from the Barbara A. Mikulski Archive for Space Telescopes (MAST) and include the latest CTE corrections for WFC3/UVIS \citep{Anderson:2021} and the improved WFC3/IR Blob Flats \citep{Olszewski:2021}. The images were aligned and drizzled using the \textsc{TweakReg} (v1.4.7) and \textsc{AstroDrizzle} (v3.1.8) tools within \textsc{DrizzlePac} (v3.1.8) \citep{Avila15}. The absolute astrometry was determined using point sources in common between the temporally drizzled images and the \href{https://www.cosmos.esa.int/web/gaia/early-data-release-3}{GAIA~EDR3} catalog \citep{GaiaCollaboration2016, GaiaDOI, GaiaCollaboration2021}. For the F475X data, we used 12 sources in common and obtained an astrometric uncertainty of 5.5~mas. Similarly, we used 19 sources for the F160W data with a resulting astrometric tie uncertainty of 19~mas. We then created a final image mosaic including cosmic ray removal and sky subtraction with north up and drizzle parameters of \texttt{pixfrac}$=0.8$ and \texttt{pixscale}$=0.03''$. The final {\it HST} images centered on the host galaxy are shown in Figure~\ref{fig:optical} with C-band radio contours overlaid. The quasi inertial reference frames defined in the radio (ICRF3) and \textit{HST} (Gaia-CRF3) images agree at the level of $\sim$10 microarcseconds \citep{Klioner22}, and hence reference frame tie uncertainties are a negligible contribution to the overall astrometric error budget, compared to the measurement of the radio and optical positions within those frames.

We used \texttt{Source Extractor} \citep{SExtractor} to determine the host galaxy center from the F160W image, and find RA(J2000)=\ra{5}{08}{03.493}$\pm$0.0116\arcsec, Dec(J2000)=$+$\dec{26}{03}{37.768}$\pm$0.0148\arcsec. Using the EVN FRB position from \cite{Nimmo22} and including the positional and astrometric uncertainties, we determine an FRB offset of $0.762 \pm 0.019$\arcsec~from the host center (see Section~\ref{sec:res} for more details). We further describe modeling of the surface brightness profile in the {\it HST} images to uncover sub-structure in Section~\ref{sec:localenv}.

\begin{figure*}
    \hspace*{-1.0cm}
    \includegraphics[width=0.47\textwidth]{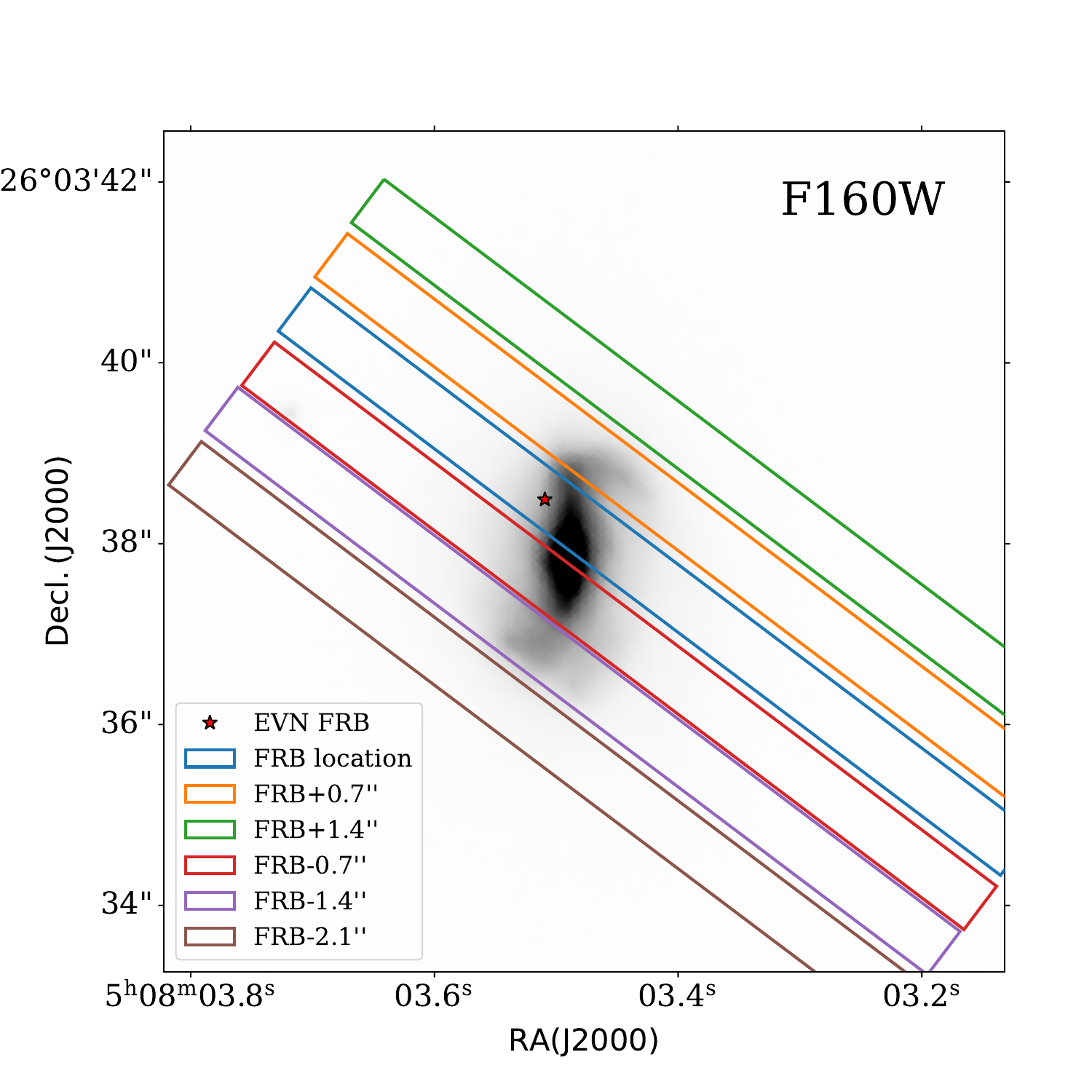}
    \hspace*{-0.5cm}
    \includegraphics[width=0.65\textwidth, trim={3cm 0 0 0}, clip]{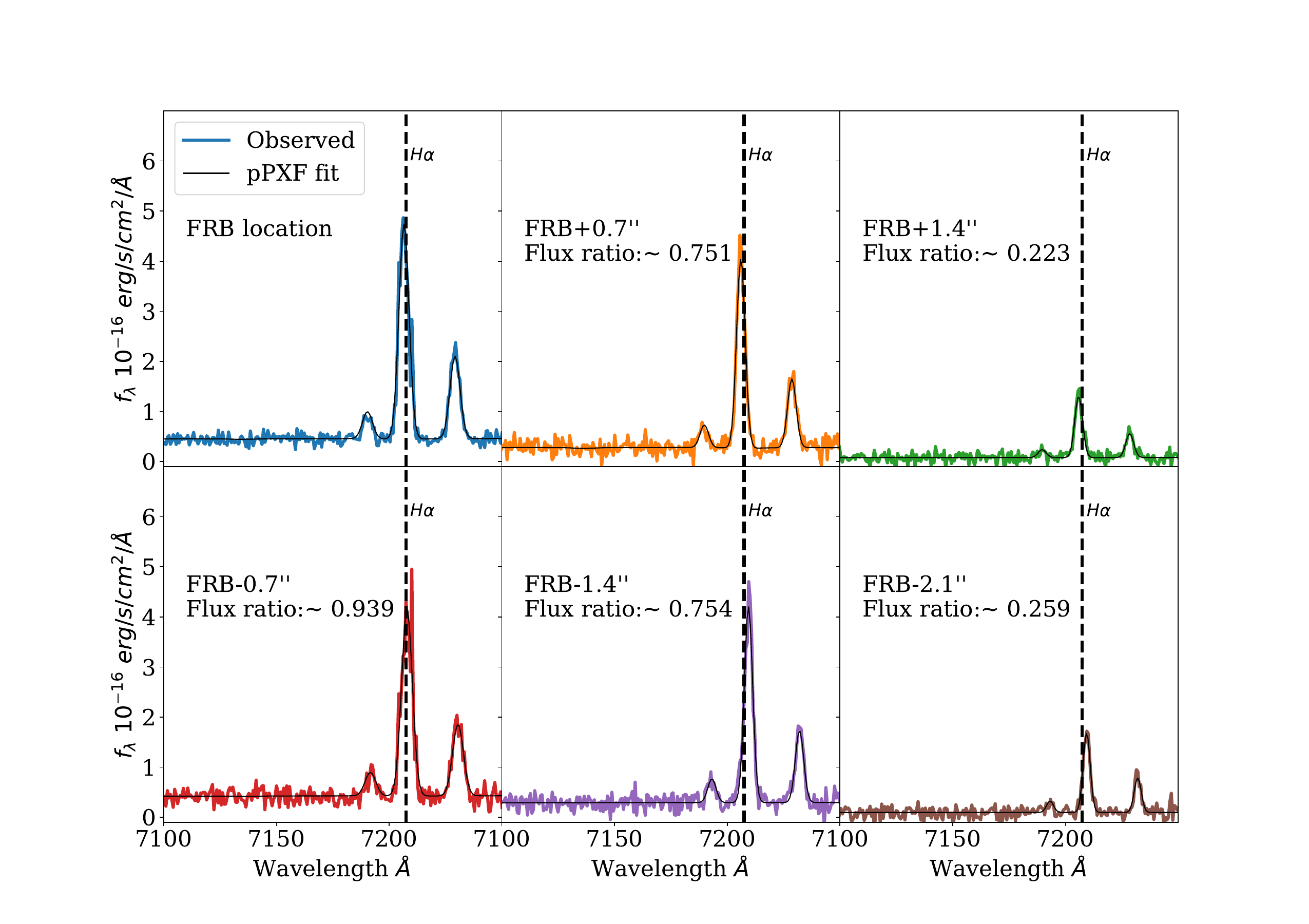} 
    \caption{\textit{Left}: Slit positions of the DEIMOS observations overlaid on the HST F160W image of the host galaxy. Each slit has a width of $0.7\arcsec$ although shown in this figure with $0.6\arcsec$ widths for clarity. The centers of each adjacent position are spaced $0.7\arcsec$ apart, thus providing full spatial coverage of the galactic disk. The blue rectangle denotes the slit position and angle aligned with the FRB position, while the remaining positions cover the rest of the galaxy.
    \textit{Right}: Spectra from the six slit positions zoomed in on $H\alpha$. The six pointings are ordered and color-coded as in the left panel. Pointings are labeled by their offsets perpendicular to the initial slit position. We fit the spectra using the pPXF routine and overlaid the best fit as the solid black curves. The vertical dashed lines show the expected location of $H\alpha$ based on the galaxy redshift. The flux ratio in each panel corresponds to the pPXF estimates of the $H\alpha$ flux values with respect to the blue pointing, i.e. the flux ratio of the blue pointing is 1. We find no enhancement in the flux ratio through the blue pointing that contains the FRB location.}
    \label{fig:deimos}
\end{figure*}

\subsection{Ground-based Spectroscopy}\label{sec:spectra}

To compare the emission line flux at the FRB location relative to the host galaxy, we observed the host of FRB\,20201124A with the DEep Imaging Multi-Object Spectrograph (DEIMOS) spectrograph mounted on the Keck II telescope on 2022 October 28 UTC (Program U129; PI: J.X.~Prochaska). We used the 600ZD grating at a central wavelength of 7000\AA. We designed a slitmask with a $0.7\arcsec$ slit to pass through the FRB location. After mask alignment and exposure, we offset the telescope along the width of the slit by $\approx 0.7\arcsec$, maintaining the same position angle to sample most of the host galaxy. In total, we exposed for 400~sec each at six different positions across the galaxy. The pointings are displayed in Figure~\ref{fig:deimos} and overlaid on the \textit{HST} F160W image of the host galaxy.

We reduced the data with the PypeIt software suite \citep[v1.12;][]{PypeIt}. PypeIt performs bias-subtraction, flat-fielding, cosmic ray masking, and wavelength-calibration of the raw frames. After the initial processing to generate calibrated 2D spectral images, the pipeline extracts 1D spectra. We subtracted the bias and flat-field corrected the raw frames using calibration frames collected in the afternoon of the observing run. The spectra were wavelength calibrated against arc spectra also collected that afternoon. We used Hg, Cd, Zn, Ne and Ar lamps for the arcs. We then flux calibrated the spectra against the archived sensitivity function available with PypeIt as opposed to generating a sensitivity function from our own standard star observations. This is because the wavelength range covered by the slit in our mask does not match that of a longslit used to observe a standard star spectrum. To ensure that the $i$-band fluxes summed over all pointings match the photometric measurement from the Pan-STARRS catalog \citep{panstarrs}, we multiplied the fluxed spectra by an additional factor of 4.1. We note that the factor required to reconcile our net synthetic flux to the Pan-STARRS value depended on the photometric band, e.g. one would have to scale the spectra by a factor of 9 instead of 4.1 to reconcile fluxes in the \textit{g}-band and 5.6 in the \textit{r}-band. This is likely due to sub-optimal seeing conditions and the atmospheric dispersion-based losses as DEIMOS does not have an atmospheric dispersion corrector. Therefore, we do not attempt to correct the spectra for internal extinction based on the flux ratios of $H\alpha$ and $H\beta$ lines.

To understand if the star formation rate is markedly enhanced at the FRB location, we use the $H\alpha$ flux ratios with respect to the slit position containing the FRB location as a proxy (right panels in Figure \ref{fig:deimos}). We find that the flux ratios decrease as the slit positions move away from the host center, as expected. We also do not find any indication of enhanced line emission at the FRB location in the 2D spectrum of the central pointing (blue slit in Fig \ref{fig:deimos}).

\begin{figure*}
    \centering
    \includegraphics[width=\textwidth]{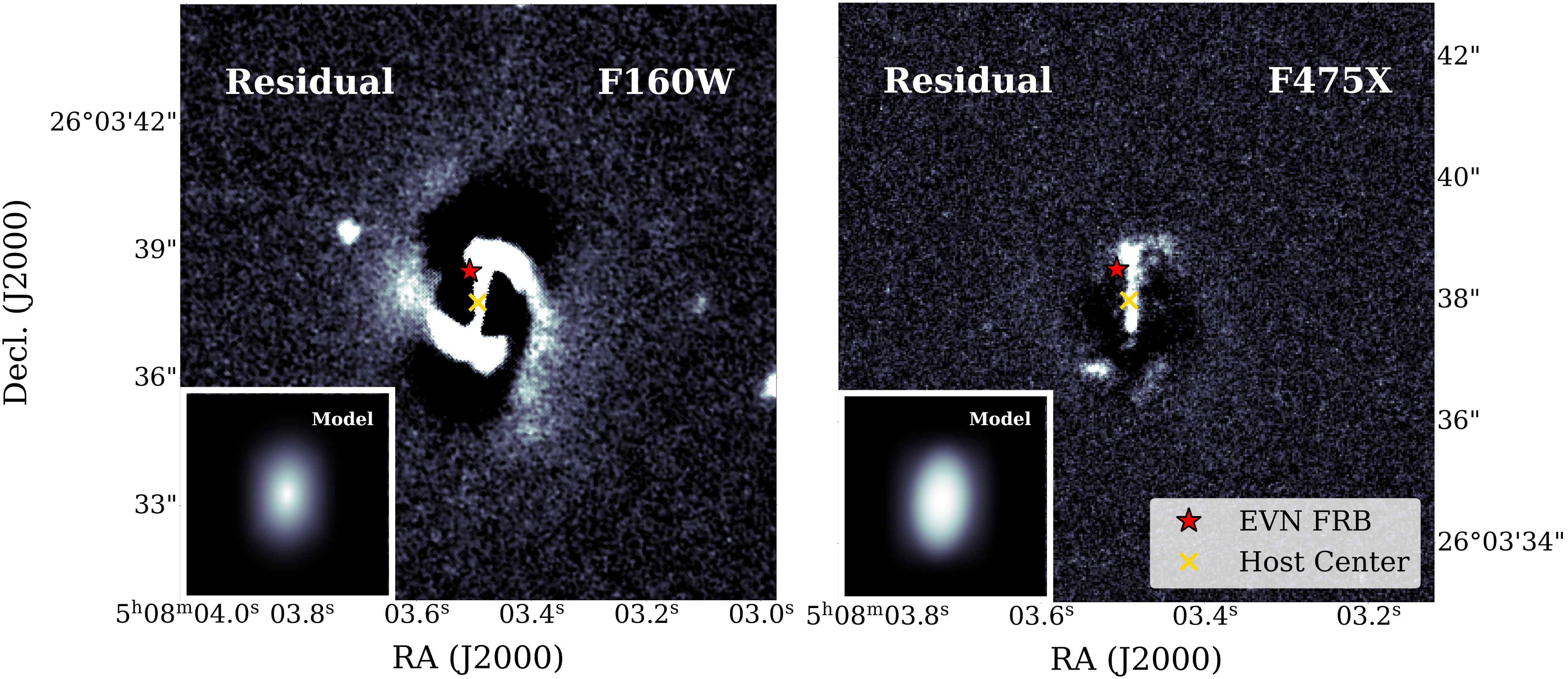}
    \caption{Residual images of the host galaxy from \texttt{GALFIT} in the F160W (14\arcsec $\times$ 14\arcsec) and F475X (10\arcsec $\times$ 10\arcsec) \textit{HST} filters. The insets are the model images from \texttt{GALFIT} with the same sizes which demonstrate that the smooth galaxy light component has been subtracted. The FRB localization and host center are marked as a red star and a yellow cross, respectively. The F160W residual image highlights the faintness of the spiral arms compared to the bar while the F475X residual image shows apparent star formation along the bar. The FRB appears to be offset to the East of the bar and not coincident with either spiral arm.}
    \label{fig:residual}
\end{figure*}

In addition, we identify faint H$\beta$ and [O\,{\sc iii}]$\lambda$4959 and $\lambda$5007 features from the slit covering the FRB location corresponding to a redshift of $z = 0.5532$. This result confirms the presence of a galaxy background to the FRB as first noted in \citet{Fong21} despite no detection in deep imaging observations \citep{Xu22}. In the 2D spectrum, the centroid of the spectral lines are $\sim$~3 pixels away from the center of the FRB host continuum trace. Assuming we are sampling the center of the background galaxy, we derive a rough position of R.A.=\ra{05}{08}{03.46}, Dec=$+$\dec{26}{03}{38.05}, consistent with the location determined in \cite{Xu22}. This corresponds to an offset of $0.8\arcsec$ southwest ($\rm PA = 235\deg$) from the FRB location. We employ PATH (Probabilistic Association of Transients to their Hosts; \citealt{PATH}) to determine the posterior probability of association assuming r = 24~mag which corresponds to a 1.06 $\times 10^{29}$\,erg\,s$^{-1}$\,Hz$^{-1}$ luminosity galaxy at $z = 0.5532$, a 10$\%$ probability that the background galaxy is unseen, an angular size of 0.5\arcsec, and a directional offset of 0.2\arcsec\ from the FRB location. We calculate a $<1\%$ posterior probability of association for the background galaxy, thus strongly disfavoring an FRB association with the background galaxy.

\section{Analysis and Results} \label{sec:res}
\subsection{Host Morphology and Star Formation}\label{sec:globalenv}

The host of FRB\,20201124A is a moderately massive star-forming spiral galaxy at $z=0.0979$ \citep{Fong21, Xu22}. A fit to the broadband spectral energy distribution (SED) of the galaxy indicates that it has a stellar population age comparable to other repeating FRBs ($\sim$2--6 Gyr; \citealt{Gordon2023}) with moderate intrinsic dust-extinction ($\sim$1--1.5 mag; \citealt{Fong21, Piro21, Ravi22}). The stellar population properties, including the star formation rate (SFR), stellar mass, and metallicity, are otherwise consistent with the known population of star-forming FRB hosts \citep{Heintz20, Bhandari22c, Gordon2023}. While many well-localized FRBs are found spatially coincident with or near the spiral arms \citep{Bassa2017, Tendulkar2017, Chittidi2021, Mannings21}, our \textit{HST} images show that FRB\,20201124A is located in the disk of its host but not directly on the main pair of spiral arm features, nor on the central bar (Figure~\ref{fig:optical}). This is consistent with the finding from Keck observations \citep{Xu22}. Moreover, FRB\,20201124A is one of only three known FRBs with clear bar structures observed in the host galaxy (along with the host of FRB\,20190608B and FRB\,20220319D, although the bar feature in both cases is less apparent with available imaging; \citealt{Chittidi2021, Ravi2023}).

To reveal the fainter underlying structures in the host, we employ \texttt{GALFIT} \citep{GALFIT} and a PSF model from TinyTim \citep{Tinytim}. TinyTim is a program that generates \textit{HST} PSFs with a fine-sampling factor of 2 relative to the image to fit the light profile of the host galaxy. For each filter, we fit the {\it HST} image with a single S\'{e}rsic profile representing the 2D surface brightness. The free parameters in the model include the centroid of the galaxy, integrated magnitude, the position angle of the major axis, ellipticity, effective radius and the S\'{e}rsic index. We show the residual images from our best fit for both filters in Figure~\ref{fig:residual}.

The most prominent morphological feature is the bar, which extends out from the nucleus and is brighter in both the optical and IR bands relative to the spiral arms. At $z\approx 0.1$, bars are a common feature among spiral galaxies \citep{Sheth2008, Melvin2014}, and most barred spiral galaxies have been observed with dust lanes \citep{Pease1917, Sandage1961} that likely contribute to obscuration of star formation. The pair of continuous spiral arms are more apparent in the rest-frame IR band and wrap around the galaxy further than exhibited by ground-based imaging (\citealt{Xu22}; Figure~\ref{fig:residual}). However, only patches of apparent star formation along the arms are visible in the optical band. Based on these features, we identify the host as an SBb, late- to intermediate-type barred spiral galaxy. In such galaxies, the bar can facilitate gas inflow and transport material from the disk into the galactic center, thereby enhancing star formation \citep{Knapen10, Wang2012}. Indeed, the radio emission tightly traces the bar and encompasses the spiral arms (Figure~\ref{fig:optical}) as expected for regions of enhanced recent star formation. The fact that the spiral arms are not immediately apparent in the bluer optical band is indicative of obscuration and supported by the dust extinction of $A_V \sim 1-1.5$~mag as measured from SED fitting \citep{Fong21, Piro21, Ravi22}.

Based on the extended nature of the radio emission and coincidence with the IR morphology (Figure \ref{fig:optical}), we conclude that the primary mechanism powering the radio emission is star formation. In this picture, as massive stars undergo core collapse, the resulting supernova remnants accelerate high-energy electrons to relativistic speeds, producing non-thermal synchrotron radiation \citep{Condon92}. At lower frequencies (1.5~GHz), the low angular resolution precludes any discernible radio features at the FRB position. At higher frequencies (3 -- 6~GHz), the radio morphology appears to trace the bar and spiral features as probed by our \textit{HST} data. We utilize the radio emission at 6~GHz to place constraints on a PRS coincident with the FRB in Section~\ref{sec:PRS}. In contrast, at 22~GHz \citep{Piro21}, the emission is concentrated on the northern side of the host galaxy where the bar connects to a spiral arm (Figure \ref{fig:optical}). Despite this difference, the diffuse radio emission at 6 and 22~GHz are broadly consistent at the 3$\sigma$ level near the FRB location.

\begin{figure}[t]
\includegraphics[width=0.45\textwidth]{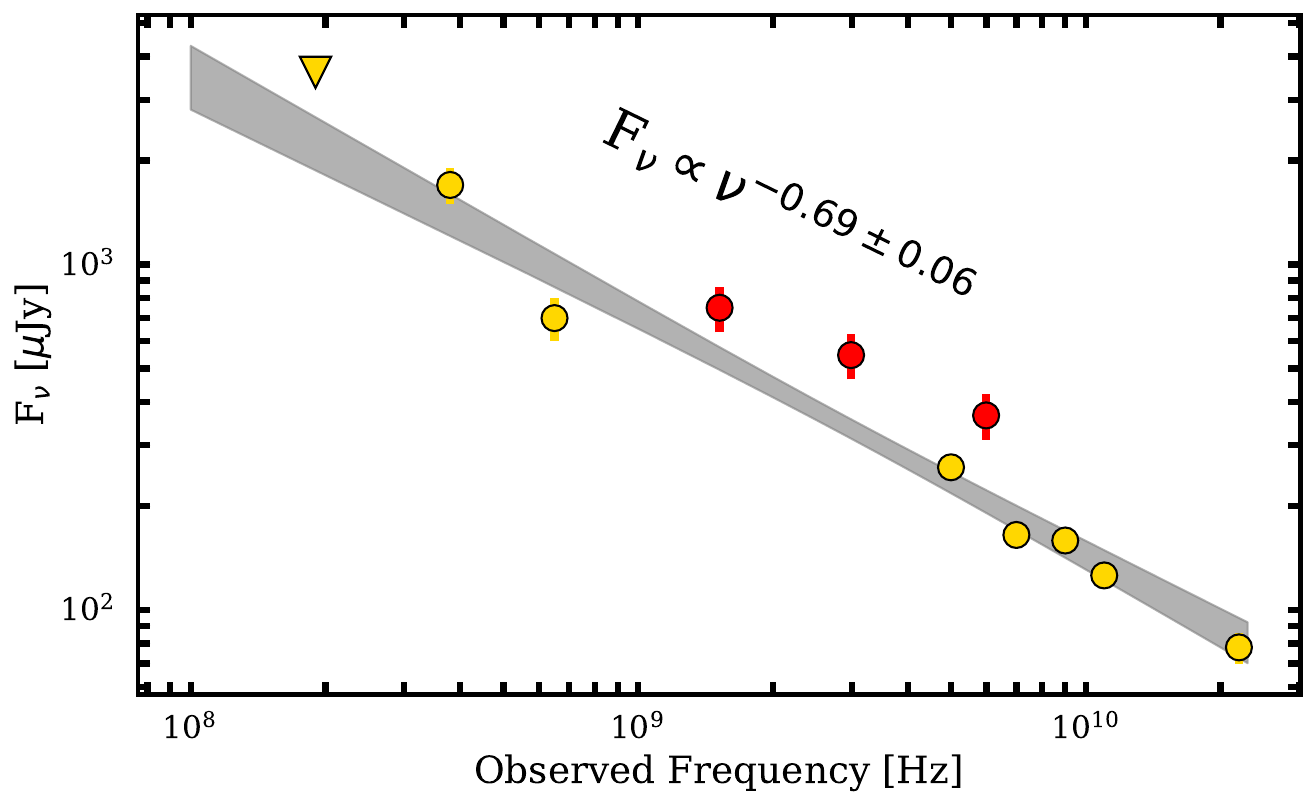}
\caption{Radio SED at 190~MHz $<$ $\nu$ $<$ 22~GHz including our data (red circles), data from the uGMRT at 380 and 650 MHz and the VLA in C- and D-configurations (yellow circles; 
\citealt{Piro21, Wharton21}). The grey shaded region corresponds to the best-fit power law (assuming F$_{\nu}$ $\propto$ $\nu^{\alpha}$) to the archival data and the 1$\sigma$ error region. We find a spectral index of $-0.69 \pm 0.06$, consistent with the expectation for star formation.}
\label{fig:SED}
\end{figure}

At GHz frequencies, the non-thermal emission from galaxies can be characterized by a power-law F$_{\nu}$ $\propto$ $\nu^{\alpha}$ with a canonical value for the spectral index $\alpha$ between $-0.7$ and $-0.75$ \citep{Gioia82, Condon92, Klein18}. In Figure \ref{fig:SED}, we plot the radio SED spanning 190 MHz -- 22~GHz. Our radio data are complemented with uGMRT measurements between 190 and 380~MHz \citep{Piro21, Wharton21} and VLA measurements between 5 and 22~GHz in C- and D-configurations \citep{Piro21}. We set the reference frequency at 3~GHz and employ \texttt{curve$\_$fit} from the \texttt{SciPy.optimize} package \citep{scipy} to fit for the spectral index assuming $F_\nu \propto \nu^{\alpha}$. We use both the literature data and our data separately due to the apparent flux density overestimation in our observations. We find that the literature data is well-characterized by a single power law with $\alpha = -0.69 \pm 0.06$ (68$\%$ confidence). We repeat the same method for our VLA data only (1.5, 3 and 6~GHz) and determine $\alpha = -0.52 \pm 0.15$. We find that although our observations produce a less precise estimate of the spectral index, it is consistent with the literature data spectral index to within 1$\sigma$. Overall, the spectral index aligns with results from previous studies \citep{Fong21, Piro21} and the expected value for star formation. 

Finally, we remark on the likelihood of any significant flux contamination from the background galaxy as discussed in Section \ref{sec:spectra}. If the background galaxy hosts a core-dominated active galactic nucleus (AGN; e.g. a blazar), we would expect a flat spectrum (i.e., a constant spectral index over a wide range of frequencies, \citealt{Urry95, Sabater19}). However, this is not the case from our SED fitting. If instead there is radio emission from star formation, the required SFR is a high value of $\approx 70\,M_{\odot}$~yr$^{-1}$ \citep{Greiner16} to even contribute up to 10\% of the observed flux. As we do not observe any separate radio and IR emission at the location of the background galaxy, we conclude that it is improbable that it contributes significantly to the radio flux density. 

\subsection{Offset and Properties at Burst Location}
\label{sec:localenv}

We leverage our \textit{HST} images to measure the precise projected offset of the FRB with respect to the host galaxy center in angular, physical, and host-normalized units. We adopt the method from \cite{Mannings21} for the offset and find that the FRB is located at an angular offset of $762 \pm 19$ mas from the host optical center in the rest-frame IR image, corresponding to a projected physical offset of $1.426 \pm 0.036$ kpc which agrees with previous measurements \citep{Fong21,Nimmo22}. 
When compared to the sample of localized FRBs, the offset of FRB\,20201124A is smaller than the median projected offset of 5.7\,kpc \citep{Bhandari22c}. However, given the range of FRB host sizes, the projected offset is less meaningful than the host-normalized offset. Thus, we use the \texttt{GALFIT}-derived value for the host effective radius of $r_{\rm e}=1.885 \pm 0.004$\,kpc (Section \ref{sec:globalenv}) and measure a host-normalized offset of $r_{\rm norm} = 0.756 \pm 0.019 \,r_{\rm e}$. This is within the offset range (0.5 -- 2.4$\,r_{\rm e}$) found by \cite{Mannings21} for a sample of FRB hosts with \textit{HST} imaging but lower than the median (0.9$\,r_{\rm e}$). Overall, FRB\,20201124A is slightly closer to its host center than average FRBs. 

Next, we calculate the fractional flux ($F_{\rm F}$) to determine how the FRB location relates to the host stellar mass and star formation by comparing the brightness at the FRB location relative to the host rest-frame optical and IR light distributions (e.g., \citealt{Fruchter06}). Specifically, it is the fraction of total flux from the host galaxy that is fainter than or equal to the flux at the FRB location, in which $F_{\rm F}=1$ indicates that the burst occurred on the brightest region in the galaxy. This method has the advantage of being independent of the host morphology and size compared to the offset measurements. We employ the method from \cite{Mannings21} and determine the flux within a 3$\sigma$ localization ellipse centered at the FRB position and weigh the values with a 2D Gaussian probability distribution. From our imaging, we measure $F_{\rm F, opt}=0.65 \pm 0.06$ and $F_{\rm F, IR}=0.46 \pm 0.02$. To first order, the FRB location is an average light location compared to the total distribution in star formation and stellar mass. This is corroborated by our DEIMOS spectra in which the $H{\alpha}$ flux ratios at and near the burst site are very similar. Thus, while the FRB is offset from regions of the strongest apparent (in the optical) star formation, it is still well within the stellar mass distribution of the host. 

While it is clear that there is some radio emission indicative of star formation at the FRB position, we can also use our {\it HST} F475X surface brightness distribution to determine if the level of star formation at the FRB position is enhanced relative to the rest of the galaxy. To this end, we compare the local flux at the FRB location with the global value from the F475X image as the rest-frame optical is typically dominated by the most recent burst of star formation. We determine the total flux within a number of 3-pixel radius apertures, weighted by the location within the localization ellipse (similar to our calculations of offset and fractional flux). We then divide by the aperture areas to obtain a surface density. We find $\rm F475X_{FRB}$ = $38.70 \pm 0.19$ \rm counts s$^{-1}$ arcsec$^{-2}$. 

Similarly, we use the effective radius of the host galaxy to calculate the total area and obtain the global $\rm F475X_{host}$ = $34.32 \pm 3.31$ \rm counts s$^{-1}$ arcsec$^{-2}$. Taking the ratio of the local and global F475X flux yields $\sim$1, suggesting that the local SFR is similar to the global mean value. Assuming there is not a significant patch of dust at or along our line of sight toward the FRB location, we can use flux in the F475X filter as a proxy for SFR. This suggests that the FRB is not in a region of significant star formation enhancement in the host, as supported by its location with respect to galaxy sub-structure. This finding is similar to the majority of other FRBs with {\it HST} imaging (with the exception of FRB\,20121102A; \citealt{Mannings21}). However, it is worth noting that these other FRBs are found to be spatially coincident with spiral arms in their respective hosts, while FRB\,20201124A is located offset from regions of the strongest star formation.

\input{tables/sfr.tex}

\subsection{Constraints on Obscured Star Formation}

While optical data are subject to the effects of interstellar dust, radio observations are often used as a probe of unobscured star formation, especially in dust-obscured galaxies such as the host of FRB\,20201124A. Thus, we calculate a radio-inferred SFR at various frequency bands using the relation from \cite{Greiner16} which extrapolates the 1.4 GHz SFR from \citet{Murphy11} assuming the same power law function form (F$_{\nu}$ $\propto$ $\nu^{\alpha}$) and a $k$-correction:

\begin{equation}
\begin{split}
 \left( \frac{\rm SFR_{radio}}{M_{\odot}\,{\rm yr}^{-1}} \right) = 0.059 \left( \frac{F_\nu}{\rm \mu Jy} \right) \left( \frac{d_{\rm L}}{\rm Gpc} \right)^2 \\ 
\times \left(\frac{\nu}{\rm GHz}\right)^{-\alpha}(1+z)^{-(\alpha +1)}
 \end{split}
\label{eqn:radioSFR}
\end{equation}

\noindent{where $F_{\nu}$ is the observed radio flux density, $d_{\rm L}$ is the luminosity distance, $\alpha$ is the spectral slope of the SED, and $z$ is the redshift of the FRB. For FRB\,20201124A, instead of using the observed flux density from our observations which suffers from a systematic offset, we derive the expected flux density at 1.5 GHz from a fit to the literature data (Section \ref{sec:globalenv}). This yields SFR$=8.9 \pm 0.7$ $\rm M_{\odot}~{\rm yr^{-1}}$, in agreement with previous radio-inferred SFRs that show slightly elevated values compared to the optical (Table~\ref{tab:sfr}).\footnote{We note that the SFR estimated using our observed flux density at 1.5~GHz is $12.31 \pm 1.87$ $M_{\odot}~{\rm yr^{-1}}$, which is consistent with the value derived from SED fitting within 2$\sigma$ and is a factor of 6 higher than the optically-inferred SFR. However, given the observed systematic offset for our data, we adopt a more conservative approach here which still points to obscured star formation.} The larger inferred SFR from the radio band relative to that inferred from the optical ($\rm SFR_{\rm opt}=3.4$ $\rm M_{\odot}~{\rm yr^{-1}}$) by a factor of $\approx$~2.5 demonstrates that the optical emission is obscured by dust. It is worth noting that different wavelengths are probing different emission mechanisms albeit similar timescales corresponding to recent star formation (i.e., synchrotron emission from SNe in the radio and young O/B stars in the optical; \citealt{Calzetti13}).  To place these measurements in the context of those from the literature, we collate existing global SFR values for the host of FRB\,20201124A from broad-band SED fits and multiwavelength observations in Table \ref{tab:sfr}. The apparent trend of increasing inferred SFRs from optical to IR to radio is a hallmark of dust attenuation at shorter wavelengths, leading to an underestimate of the true SFR at optical (and UV) wavelengths.}

\begin{figure}[t!]
\includegraphics[width=0.51\textwidth]{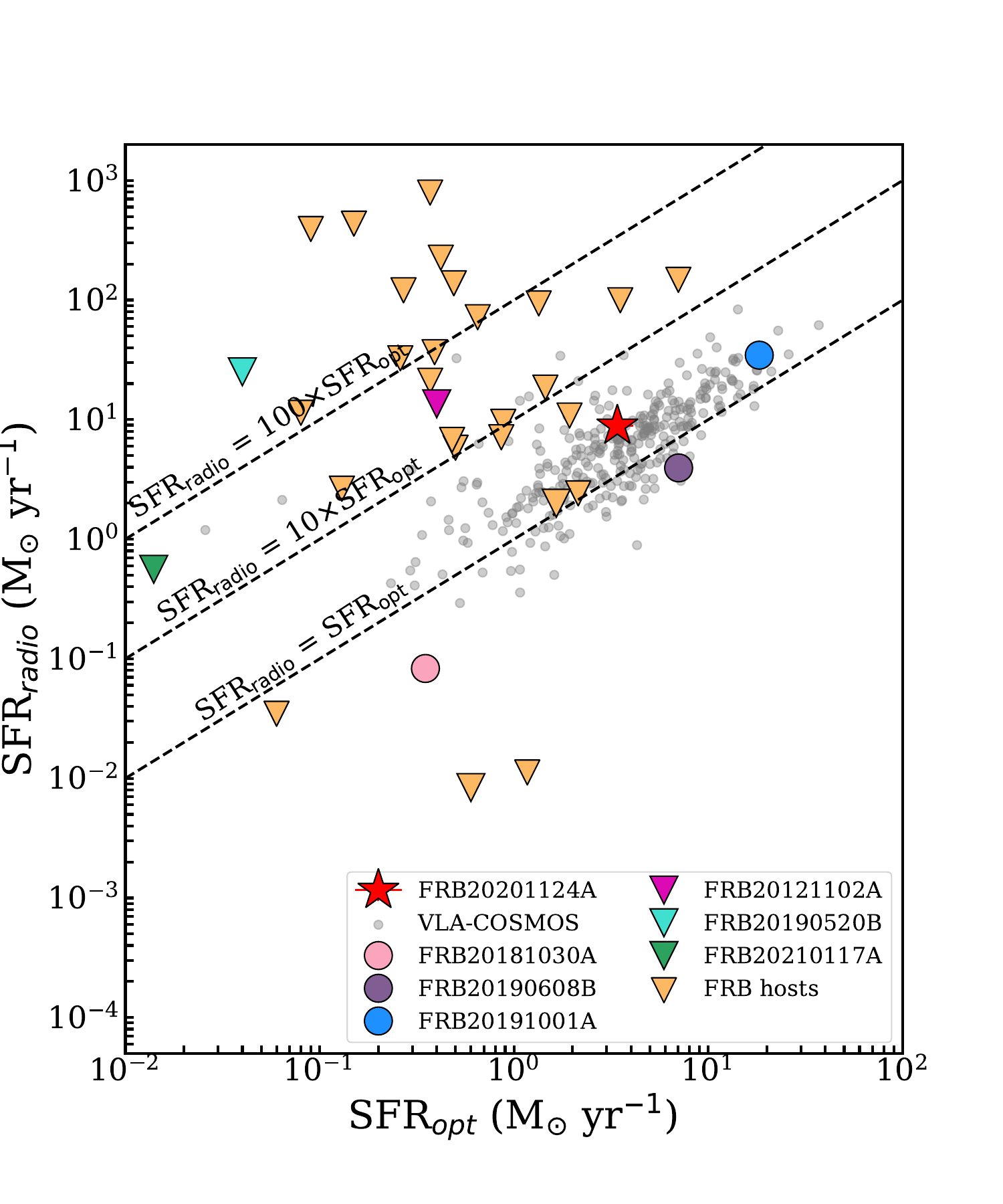}
\caption{Radio versus optical-inferred SFRs for FRB\,20201124A (red star), FRB\,20181030A (pink circle; \citealt{Bhardwaj21}), FRB\,20190608B (purple circle; \citealt{Bhandari20a}), and FRB\,20191001A (blue circle; \citealt{Bhandari20b}). Triangles denote 3$\sigma$ upper limits for FRB\,20121102A (magenta; \citealt{Marcote17}), FRB\,20190520B (turquoise; \citealt{Bhandari23b}), and FRB\,20210117A (green; \citealt{Bhandari23}) and localized FRBs with existing radio limits \citep{Law22, Law23}. Shown for comparison is a sample of radio-selected star-forming galaxies at z $\lesssim$ 0.3 from the VLA-COSMOS Source Catalog \citep{Smolvic17}. Dashed lines indicate SFR$_{\mathrm{radio}}$ = SFR$_{\mathrm{opt}}$, $10 \times$SFR$_{\mathrm{opt}}$, and $100\times$SFR$_{\mathrm{opt}}$. FRB\,20201124A is the FRB host with the most dust-obscured star formation.}
\label{fig:SFRs}
\end{figure}

To place the level of obscured star formation in context with other FRB hosts and the general galaxy population, we show the radio-inferred SFR versus optically-derived SFRs (typically from H$\alpha$ measurements or SED modeling) for a variety of sources in Figure~\ref{fig:SFRs}. For FRBs, in addition to FRB\,20201124A, we include the hosts of FRB\,20181030A \citep{Bhardwaj21}, FRB\,20190608B \citep{Bhandari20a}, and FRB\,20191001A \citep{Bhandari20b} which are the only known FRB hosts with radio emission due to star formation. We use Equation~\ref{eqn:radioSFR} to convert the published radio fluxes to radio-inferred SFRs assuming a canonical value of $\alpha = -0.7$ and the FRB redshifts. In the case of FRB\,20191001A, we instead use the published spectral index of $\alpha = -0.8$ \citep{Bhandari20b}.\footnote{The SFR derived here is $\sim$3 times higher than the published value from \cite{Bhandari20b} using a different method.} We plot the upper limits on the radio-inferred SFR from measurements of compact persistent radio emission associated with FRB\,20121102A \citep{Marcote17} and FRB\,20190520B \citep{Bhandari23b}. We also incorporate radio limit on star formation for FRB\,20210117A, the only non-repeating FRB in a dwarf host galaxy \citep{Bhandari23}. For FRB\,20200120E, which harbors a known low-luminosity AGN, we derive a conservative upper limit on the star formation-related radio emission using the observed 8.4 GHz flux density for the central AGN \citep{Miller10}. Finally, we complement our sample with 24 existing radio continuum limits from \cite{Law22, Law23}, previously used to place limits on PRS emission but which we now use as limits on star formation in the FRB hosts.

For the optically-inferred SFRs, we adopt the $\mathrm{SFR_{opt}}$ values from \cite{Heintz20} which are derived from $\mathrm{H{\alpha}}$ line luminosities. We note that the $\mathrm{H{\alpha}}$-inferred SFR for FRB\,20121102A is obtained from \cite{Tendulkar2017}. For cases in which the $\mathrm{H{\alpha}}$-inferred SFR is not available, we adopt SFRs obtained from broad-band SED fitting \citep{Bhandari20a, Hatsukade22, Bhandari23, Gordon2023, Law23}. We note that in the case of the Galactic FRB\,20200428A, we use the SFR derived from the Cosmic Background Explorer (COBE) observations of N\,{\sc ii} 205$\mu$m emission \citep{MW97} as reported in \citet{Chomiuk11}. Thus, our literature sample of 32 bursts comprises all FRB hosts with relevant radio limits or detections and accompanying optically-inferred SFRs. The results and references are listed in Table~\ref{tab:all_sfr}. For comparison, we include a sample of star-forming field galaxies that are radio-selected at $z \lesssim 0.3$ from the VLA-COSMOS Source Catalog \citep{Smolvic17}. 

\input tables/radio_opt_SFR.tex

Figure~\ref{fig:SFRs} shows that out of the four FRB hosts with detectable radio emission attributable to star formation, FRB\,20201124A is one of two hosts with evidence for dust-obscured star formation and the most heavily dust-obscured with $\mathrm{SFR_{\rm radio}} \approx 2.5 \times \mathrm{SFR_{\rm optical}}$. Furthermore, FRB\,20201124A is well above the majority of other normal star-forming galaxies at similar redshifts (e.g., the VLA-COSMOS sources). A comparison of the radio and optical SFRs suggests that FRB\,20190608B does not have obscured star formation, commensurate with the low $A_V$ value \citep{Gordon2023} whereas FRB\,20191001A shows some mild level of dust obscuration. In contrast, the upper limits for FRB\,20121102A, FRB\,20190520B, and FRB\,20210117A are not constraining enough to place a meaningful limit on obscuration.  However, it is worth mentioning that the known PRSs are also too luminous (L$_{PRS}$ $>$ 10$^{29}$ ergs$^{-1}$Hz$^{-1}$; \citealt{Law22}) and compact in size ($<$ 1pc; \citealt{Marcote17}) to be caused by ongoing star formation. Moreover, deep radio upper limits from \citet{Law22, Law23} can rule out dust obscuration at a level similar to FRB\,20201124A for five FRBs. However, for the majority of FRB hosts, deeper radio observations are required to place tighter constraints on the existence or absence of obscured star formation.

\subsection{Constraints on a Compact Persistent Radio Source}
\label{sec:PRS}

Taking advantage of our high-resolution radio observations, we also search for a PRS coincident with the FRB location using our VLA C-band observations. The bulk of the radio emission has a complex morphology that is clearly linked to star formation. Indeed, while Figure \ref{fig:optical} shows there is no clear detection of a point-like source at the location of the FRB that would indicate a PRS similar to those associated with FRB\,20121102A and FRB\,20190520B \citep{Chatterjee17,Niu22}, the morphologically complicated host radio emission makes the process of placing upper limits on any such compact PRS challenging. To do so, we first estimate the peak emission at the burst site, and then attempt to subtract a conservative estimate for the background SFR-related radio emission.

    At the FRB location, the measured 6~GHz radio brightness is $\approx$ 15\,$\mu$Jy/beam. However, based on our radio and optical morphology, we consider it plausible that the entirety of this emission is due to star formation. Figure~\ref{fig:radio} shows that the region surrounding the burst site is enclosed by emission at or above the 4$\sigma$ level (8\,$\mu$Jy/beam). If we crudely use this as an estimate for SFR emission at the FRB location, we are left with an upper limit of 7\,$\mu$Jy for any compact PRS.\footnote{Our robust weighted image (higher resolution than natural weighted images; not shown) yields a consistent result, with a brightness at the FRB location of 9 $\mu$Jy/beam on a local background of $\sim$2 $\mu$Jy/beam. This leads to a comparable upper limit of $\sim$7 $\mu$Jy for any compact PRS at the FRB location. Thus, our conclusions are insensitive to the weighting scheme.} To ensure we do not underestimate the flux density, we employ a more conservative limit of 10\,$\mu$Jy for the potential PRS. We note that previously published EVN observations with an image rms of 14 $\mu$Jy/beam \citep{Nimmo22} would therefore not detect such a PRS (our VLA observations have an RMS of $ \sim 2\,\mu$Jy/beam, a factor of seven more sensitive). Our conservative 10 $\mu$Jy limit corresponds to $L_{\nu}$ = 2.6 $\times 10^{27}$\,erg\,s$^{-1}$\,Hz$^{-1}$ (6~GHz). However, we also note that the flux densities derived in this work are $\sim$30\% higher, on average, than what is obtained from a fit to archival values, which could be due to multiple issues as described in Section \ref{sec:obs}. If our measured flux densities were scaled downwards by 30\% to match the archival values, the limit becomes even tighter at $L_{\nu}$ = 1.8 $\times 10^{27}$\,erg\,s$^{-1}$\,Hz$^{-1}$ (6~GHz). The inferred value is deeper than the existing PRS limit placed by \cite{Ravi22} of $L_{\nu}$ = 3 $\times 10^{28}$\,erg\,s$^{-1}$\,Hz$^{-1}$ (1.4~GHz) for FRB\,20201124A and among the deepest PRS limits to date in terms of luminosity (e.g., FRB\,20180916B; \citealt{Marcote20} and FRB\,20220319D; \citealt{Law23}). Notably, the inferred luminosity of the potential PRS is fainter than the two known PRSs\,20121102A and 20190520B by two orders of magnitude \cite{}. Moreover, our limit is also deeper than the median value ($L_{\nu} \lesssim$ 2.5 $\times 10^{29}$\,erg\,s$^{-1}$\,Hz$^{-1}$) obtained for upper limits on PRS emission (Table \ref{tab:all_sfr}) compiled by \cite{Law22, Law23}.

\section{Discussion} \label{sec:discuss}
\subsection{Implications for the Progenitor of FRB\,20201124A} \label{sec:progenitors}

Shortly after the initial discovery of FRB\,20201124A \citep{CHIME_201124A}, the source was observed to enter periods of extreme activity \citep{Lanman22, Xu22, Zhou22}, becoming one of the most actively repeating FRBs to date. Similar to other repeating FRBs, FRB\,20201124A exhibits a high linear polarization fraction, flat polarization angles, and downward drifts in frequency \citep{Hilmarsson21}. Strikingly, it is the first repeating FRB to show clear signs of circular polarization up to 75$\%$ \citep{Xu22}, and one of the few FRBs to exhibit strong rotation measure (RM) fluctuations \citep{Yang22,Xu22}. In addition, FRB\,20201124A now joins a small sample of four FRBs with detectable radio emission due to star formation in the host galaxy \citep{Bhandari20a, Bhandari20b, Bhardwaj21}. Previous deep ground-based imaging showed clear bar and spiral morphology within its host galaxy \citep{Xu22}, consistent with our deep \textit{HST} imaging. They find the local environment to be inconsistent with a supernova explosion \textit{in situ} due to its offset from the main locus of the galaxy, at an interarm region. 

As shown in Figure \ref{fig:residual}, the host galaxy exhibits a bulge with winding spiral arms and a strong bar that a majority of disk galaxies (up to 70$\%$ when observed in the near-IR) have at both low and high redshifts \citep{Aguerri09}. While the FRB location is not coincident with the bar, the bar does affect the distribution of star formation sites \citep{Philips96}. In the case of an SBb galaxy, star formation is more intense in a circumnuclear ring structure and the junction where the bar meets the stellar spiral arms. From the optical morphology in Figure~\ref{fig:residual}, we observe pockets where the emission is not enshrouded by dust, pointing to an inhomogeneous distribution of dust within the host, and aligned with the expected patchy dust distributions in SBb galaxies \citep{Philips96}. Thus, it is reasonable to expect star-forming regions to extend to the burst site.

Indeed, our radio observations newly uncover obscured star formation throughout the host and extending to the FRB site. Thus, the most natural scenario is that the FRB originated from a young progenitor born \textit{in situ}. In this case, the progenitor, possibly a magnetar, is formed via core collapse of a young massive star embedded in a dust cloud. Indeed, \cite{Mannings21} and \cite{Bhandari22c} find the radial distribution of FRBs are indistinguishable from those of CCSNe. Subsequent studies by \cite{Bochenek21} and \cite{Bhandari22c} also find consistencies between the host environments of FRBs and CCSNe, strengthening the case for a magnetar progenitor population born from the explosions of young massive stars. In contrast, superluminous supernovae (SLSNe) and long-duration $\gamma$-ray bursts (LGRBs) are predominantly found in the brightest regions of low-mass host galaxies \citep{Lunnan15}, a distinctly separate type of local and global environment compared to that of FRB\,20201124A. 

Other \textit{in situ} scenarios, such as the formation of a magnetar via the accretion-induced collapse (AIC) of a massive white dwarf (WD, \citealt{Margalit19}) in a close binary with either a non-degenerate companion or another WD, are also viable. Such formation channels have been proposed for FRB\,20200120E due to its globular cluster environment \citep{Bhardwaj21_M81,Kirsten22} in which old systems like WDs are more common than young massive stars. However, such an \textit{in situ} delayed channel is not necessarily required by our observations for FRB\,20201124A.

On the other hand, to explain the apparent offset from the central bar structure and spiral arms in which the majority of star formation is occurring, it is interesting to explore migratory progenitors. While the galactocentric offset of FRB\,20201124A is smaller than the median for the FRB population, it is not particularly close to any region of enhanced star formation, and thus, we explore the possibility that it could have traveled with some velocity from its birth site. In the context of a magnetar progenitor, we explore a few such formation channels. 

Within the ``migratory'' models, the most natural channel is a magnetar born from an energetic CCSN with a significant natal kick, as in the case of many Galactic radio pulsars \citep{Lyne94, Claude06}. Well-measured velocity distributions of pulsars demonstrate that NSs typically have natal kick velocities of $\sim 100-1000$ km\,s$^{-1}$ (e.g., \citealt{Hobbs05,Deller19}). We estimate the projected distance from the FRB location to the center along the bar to be $\sim$ 0.4\,kpc. Given the range of kick velocities, this would require minimum travel times of $\sim 0.5 -4$\,Myr to achieve the observed offset assuming the NS progenitor originated in the bar, far exceeding typical active magnetar lifetimes of $\lesssim 0.1$~Myr \citep{Kaspi07}. If we instead assume an edge case in which the FRB progenitor originated from the edge of the bar ($\sim$ 0.2\,kpc projected offset) and assume a maximum kick velocity of 1000 \kms\ for magnetars, we obtain a minimum age of $\sim$ 0.2 Myr, still at odds with active magnetar lifetimes.

Another possible scenario is the remnant of a runaway O or B star through dynamical ejections similar to those observed from open clusters \citep{Blaauw61, Poveda67, Fujii11, Oh16}. Approximately 1--10$\%$ of O and B stars are runaways \citep{Fujii11} and these massive young stars can create compact NS via supernova explosions, which may preferentially happen earlier in the lifetime of those clusters when their stellar densities are higher \citep{Chen04}. The peculiar velocities of runaway OB stars range between $30 - 200$ km\,s$^{-1}$ \citep{Hoogerwerf2000} with respect to the mean Galactic rotation. Assuming a main sequence age range of 3--12 Myr for runaway O stars and 12--70 Myr for runaway B stars with initial masses $>$8~$M_{\odot}$ that can produce NS \citep[assuming 8--18~$M_{\odot}$ for B stars and $>$18~$M_{\odot}$ for O stars and models in][]{choi16}, we calculate the projected tangential velocity from the edge of the bar which would place a lower limit on the velocity of the runaway systems. We find that at a minimum, a runaway O star would need to travel at a speed between $\gtrsim$ 16 -- 65 \kms\ to achieve the observed offset, well within the expected velocity for a runaway progenitor. However, due to the longer lifetime of B stars, the inferred velocity ($\gtrsim$ 3 -- 16 \kms) suggests that the star would not necessarily need to be a runaway to travel to the observed location of the FRB. Thus, if the progenitor of FRB\,20201124A is a runaway, it would most likely be a massive and young O star. But given the small fraction of runaway OB stars in general, this scenario is likely not a common channel for most FRBs based on their high occurrence rates.

We finally remark on a third scenario for a migratory progenitor: a magnetar born from a binary neutron star (BNS) merger \citep{Margalit19, Wang20}. The recent claim of an association between the BNS merger GW\,190425 and FRB\,20190425A \citep{Moroianu23} makes this an interesting formation channel to consider for FRB\,20201124A (but also see \citealt{Bhardwaj23}). If we assume a maximum systemic velocity of 240 \kms\ for a double NS system found in the Galactic disk \citep{Tauris17}, it would require 0.8 -- 1.6 Myr to travel from the edge and center of the bar respectively. This age is much smaller than the inspiral times for BNS mergers ($\sim$ tens to hundreds of Myr at minimum; \citealt{Zevin22}), thus well within the viable timescale for the production of a magnetar remnant. However, while the offset can be explained by the kick velocity and age, we find the BNS merger scenario improbable by invoking the luminosity constraint on the PRS in the following section.

\subsection{Models for a putative persistent radio source} \label{subsec:discuss_PRS}
Despite the non-detection of any compact radio source coincident with FRB\,20201124A, the constraints we have placed on the PRS are nevertheless meaningful when applied to the current models of PRSs. To this end, we first consider the magnetar nebula model which was initially invoked to explain the observed properties of PRS\,20121102A (e.g., size, luminosity; \citealt{Yang16, Beloborodov17, Dai17, Margalit18, Li-QC20}). In this picture, the PRS originates from a magnetized electron-ion nebula powered by a flaring young magnetar that can be formed through both prompt and delayed channels as discussed in Section \ref{sec:progenitors}. The difference in each scenario depends on the surrounding environment that dictates the evolution of the magnetar wind nebula.

We adopt the analytic approach as described by \cite{Margalit19} for the expanding nebula in which the nebula luminosity depends on the energy injection rate of the magnetar $\dot{E}$ and the density of the surrounding medium, parameterized by 
$M_{\rm ej} / v_{\rm ej}^3$/t$^3$ where $M_{\rm ej}$ and $v_{\rm ej}$ 
are the total mass and mean velocity of the expanding ejecta, respectively. 
In order to satisfy our PRS luminosity limit $L_{6.0} \lesssim 1.8 \times 10^{27}$\,erg\,s$^{-1}$\,Hz$^{-1}$, the energy injection rate
must satisfy:

\begin{equation}
\begin{split}
 \dot{E} < 2 \times 10^{39}\,{\rm erg\,s}^{-1}\, \chi_{0.2}^{0.72} \sigma_{-1}^{0.70} M_{{\rm ej},10}^{-0.42} v_{{\rm ej},9}^{1.26} t_{100}^{0.64}
\end{split}
\end{equation}

\noindent
where 
$\chi_{0.2} = \chi/0.2\,{\rm GeV}$, $\sigma_{-1} = \sigma/0.1$ denote the mean energy per ion ejected in the magnetar wind and the wind magnetization; $M_{{\rm ej},10} = M_{\rm ej}/10M_\odot$, $v_{{\rm ej},9} = v_{\rm ej}/10^9\,{\rm cm\,s}^{-1}$ represent the ejecta mass and velocity;
and $t_{100} = t/100~ \mathrm{yr}$ denotes the age of the expanding nebula. 
For these fiducial parameters, relevant for a magnetar born through core-collapse of a massive star, this implies an internal magnetic field strength of the putative magnetar $\lesssim 10^{16}\,{\rm G}$ (see Eq.~2 of \citealt{Margalit19}). This conclusion does not depend on the (unknown) source age if we assume that it satisfies $t < t_{\rm mag}$, such that $t_{\rm mag}$ is the magnetar's active lifetime (Eq.~1 of \citealt{Margalit19}). From Figure~4 of \citet{Margalit19}, this allows us to rule out a significant portion of the $\dot{E}$-ejecta density space that may be realized in the BNS merger scenario. While magnetars born from LGRBs, SLSNe, and AIC, which are compatible with lower energy injection rates remain plausible, the location of FRB\,20201124A in its host makes these progenitor scenarios less likely (see Section~\ref{sec:progenitors}). 

One caveat to note is that the model assumes a constant energy injection rate. While this makes the calculations analytically tractable, it was shown that a constant $\dot{E}$ is not able to explain the detailed properties of FRB\,20121102A \citep{Margalit18}. On the other hand, a $\dot{E}$ that declines as a function of time (as invoked for FRB\,20121102A) would generally lead to less luminous persistent radio emission at a fixed time $t$. This would loosen our constraints on the magnetar $B$-field, since it would become easier to satisfy our PRS upper limits. We therefore cannot definitively rule out a magnetar progenitor with an internal magnetic field $\gtrsim 10^{16}\,{\rm G}$. However, such high fields would generally suggest a short active lifetime \citep[e.g.,][]{Margalit19}, which may start to become in tension with the lack of free-free absorption at $\sim$ GHz frequencies.

An alternative scenario for FRB progenitors was introduced by \citet{Sridhar+21} in which FRB pulses are emitted along the jet axis of an accreting compact object, and the associated PRS synchrotron emission is powered by particles heated at the termination shock of the outflowing super-Eddington disk wind \citep[`Hypernebula';][]{Sridhar22, Sridhar+23}. Unlike a magnetar-powered engine, the activity rate of FRBs from an accretion-powered engine does not necessarily diminish as the system ages. In this scenario, a typical burst luminosity $L_{\rm FRB}\sim10^{40}\,{\rm erg \ s^{-1}}$ for FRB\,20201124A \citep{Xu22} would require an accreting engine with a mass transfer rate of $\dot{M}\sim10^3\,\dot{M}_{\rm Edd,10}$ where $\dot{M}_{\rm Edd,10}\equiv L_{\rm Edd,10}/(0.1c^2)$ is the Eddington rate of an accreting engine with a mass of $M_\bullet=10\,M_\odot$, and $L_{\rm Edd,10}\simeq1.5\times10^{39}{\rm erg/s}$ is the corresponding luminosity. At this $\dot{M}$, the outflowing disk winds can contribute $\sim120\,{\rm pc\,cm^{-3}}$ to the overall dispersion measure (DM) of FRB\,20201124A \citep[i.e., comparable to the host DM measurement from][]{Ravi22, Xu22} after $\sim10^5$\,yr of expansion \citep[Eq.~12 of][]{Sridhar22} for the PRS. 

Using Eq.~10 and Eq.~28 of \cite{Sridhar22}, the size of the shell formed from the interaction of the disk wind with the surrounding medium and the radio synchrotron-bright region (the ‘nebula’) are $R_{\rm sh}\ge25\,{\rm pc}$ and $R_{\rm n}\lesssim1\,{\rm pc}$, respectively. Even for the highest angular resolution C-band images (robust weighting, for which the angular resolution of 0.23\arcsec corresponds to a linear size of 430 pc), a size of this source would be effectively point-like to our VLA observations. This means that our earlier comparisons against the peak brightness at the FRB location provide valid upper limits on the PRS flux density. The bright nebula region can be detectable in the radio at a luminosity of $\sim10^{39}\,{\rm erg \ s^{-1}}$ up to $\sim10^5$\,yr. However, given the active lifetime of the system also being $\sim10^5$\,yr ($\sim M_\star/M_\bullet$, for an assumed post-main sequence companion star with a mass of $\sim30\,M_\odot$), the luminosity would fall off precipitously after $\sim10^5$\,yr, in agreement with our flux density upper limit  ($<10^{37}\,{\rm erg/s}$) that is two orders of magnitude lower. Thus, the hypernebula model is viable given the luminosity constraints from our radio observations.

Finally, multipath propagation from a magnetized plasma screen has been hypothesized to produce the observed polarization behaviors including frequency-dependent depolarization and the mean RM ($|$RM$|$) in FRBs, notably the ones associated with a PRS \citep{Beniamini21, Feng22, Yang22}. In this model, the RM scatter ($\sigma_{RM}$) that causes the depolarization at low frequencies is correlated with $|$RM$|$ and could also be proportional to the specific luminosity of the PRS (see Eq.~42 of \citealt{Yang22}). 

The $|$RM$|$ of FRB\,20201124A is observed to decrease from 889.5\,rad\,m$^{-2}$ to 365.1 \,rad\,m$^{-2}$ over a 10-day timescale \citep{Jiang22, Xu22} and $\sigma_{RM}$ is found to be 2.5 rad\,m$^{-2}$ \citep{Feng22}. We choose FRB\,20190520B as our reference case since its RM is closer to FRB\,20201124A than that of FRB\,20121102A and scale the $\sigma_{RM}$ with PRS\,20190520B. We find that the PRS luminosity is in agreement with the upper limit of $L_{\rm 6.0 \ GHz}$ $\lesssim$ 1.8 $\times 10^{27}$\,erg\,s$^{-1}$\,Hz$^{-1}$ based on our VLA observations within a factor of 2. Consistency with our upper limits for the PRS luminosity could be maintained by, for instance, reducing the fraction of synchrotron-radiating electrons in the GHz band ($\zeta_e$) by a factor of 2. It is worth noting that if we chose PRS\,20121102A instead, the PRS luminosity would be higher than our derived upper limit by roughly an order of magnitude. However, given there are other free parameters such as the screen radius (R) and thermal Lorentz factor ($\gamma_{\mathrm{th}}$) to adjust in the equation, this model remains plausible.

\section{Conclusions} \label{sec:conclude}
We have presented high-resolution VLA and \textit{HST} observations, as well as Keck spectroscopy of the host galaxy of the extremely active repeating FRB\,20201124A. These observations map the morphology of star formation throughout the host and near the location of the FRB. We also performed a deep search for a potential PRS coincident with FRB\,20201124A. Our main results are as follows:

\begin{itemize}
\item The radio emission at 1.5 and 3~GHz is marginally resolved and centered on the host galaxy. The higher spatial resolution afforded by our 6~GHz observations show that the radio emission is extended, and tightly traces the IR morphology of the host galaxy, especially along the bar and enclosing the pair of spiral arms. Moreover, the radio emission extends to the position of the FRB indicating some amount of star formation at the FRB location.

\item Our \textit{HST} observations clearly display the bar and spiral arm features of the host in the rest-frame IR and indicate a patchy dust distribution in the optical band. A comparison of the local and global values of fractional flux and SFR density indicate that the FRB\,20201124A location is average in both its stellar mass and star formation distributions.

\item The host galaxy of FRB\,20201124A has a radio-inferred SFR of $8.9 \pm 0.7$ $M_{\odot}~{\rm yr^{-1}}$, a factor of $\approx 2.5$ higher than the SFR inferred from H$\alpha$ luminosity, demonstrating the presence of dust-obscured star formation throughout the host environment. 

\item A comparison of all four FRB hosts with radio emission powered by star formation, as well as all 29 additional limits on radio star formation in FRB host galaxies, demonstrates that FRB\,20201124A is one of two known FRB hosts with dust-obscured star formation and the most dust-obscured to date.

\item Deep radio imaging does not reveal a compact PRS co-located with FRB\,20201124A, thereby constraining the luminosity of the PRS to be $L_{\rm 6.0 \ GHz}$ $\lesssim$ 1.8 $\times 10^{27}$\,erg\,s$^{-1}$\,Hz$^{-1}$. This is two orders of magnitude lower than the two known PRSs associated with FRBs. Our limit is consistent with both the magnetar wind nebula model assuming a constant energy injection rate and the hypernebula model assuming a PRS age of $\gtrsim 10^{5}$ yr.

\item The fact that the radio emission due to star formation extends to the burst site suggests that the progenitor of FRB\,20201124A was born \textit{in situ}, for example, a magnetar born from the core-collapse of a massive star. Migratory progenitor scenarios such as runaway massive stars, although plausible, seem less likely as they are not necessary to explain the observed FRB offset from the bar. 
 
\end{itemize}

In this study, we highlight the importance of utilizing joint rest-frame optical and radio observations of FRB host galaxies. In particular, multiwavelength follow-up of FRB sources provides a more comprehensive view on the environments surrounding these events and ultimately their progenitor channel(s). The extremely active nature of FRB\,20201124A, combined with its apparent offset from any active regions of star formation in the optical bands, sets a precedent for future detections of FRBs as it offers a discerning view on the current possible progenitor scenarios. Future multiwavelength studies of FRB hosts and their morphological properties, both globally and local to the FRB environment, will lend critical insight into the formation channel(s) of FRBs.

\section{Acknowledgments}
We thank Drew Medlin from the NRAO for the reduction of our radio data, and Joseph Michail and Michael Zevin for helpful discussions. Y.D. is supported by the National Science Foundation Graduate Research Fellowship under Grant No. DGE-1842165. T.E. is supported by NASA through the NASA Hubble Fellowship grant HST-HF2-51504.001-A awarded by the Space Telescope Science Institute, which is operated by the Association of Universities for Research in Astronomy, Inc., for NASA, under contract NAS5-26555. S.B. is supported by the Dutch Research Council (NWO) Veni Fellowship (VI.Veni.212.058). W.F. gratefully acknowledges support by the David and Lucile Packard Foundation, the Alfred P. Sloan Foundation, and the Research Corporation for Science Advancement through Cottrell Scholar Award \#28284. C.D.K. is partly supported by a CIERA postdoctoral fellowship. B.M. acknowledges financial support from the State Agency for Research of the Spanish Ministry of Science and Innovation under grant PID2019-105510GB-C31/AEI/10.13039/501100011033 and through the Unit of Excellence Mar\'ia de Maeztu 2020--2023 award to the Institute of Cosmos Sciences (CEX2019-000918-M). N.S. acknowledges the support from NASA (grant number 80NSSC22K0332), NASA FINESST (grant number 80NSSC22K1597), and Columbia University Dean's fellowship. Y.D., T.E., W.F., A.C.G., C.D.K., A.G.M., J.X.P., S.S., and N.T. acknowledge support from NSF grants AST1911140, AST-1910471 and AST-2206490 as members of the Fast and Fortunate for FRB Follow-up team. The Fong Group at Northwestern acknowledges support by the National Science Foundation under grant Nos. AST-1814782, AST-1909358 and CAREER grant No. AST-2047919.

The National Radio Astronomy Observatory is a facility of the National Science Foundation operated under cooperative agreement by Associated Universities, Inc.

This research is based on observations made with the NASA/ESA Hubble Space Telescope obtained from the Space Telescope Science Institute, which is operated by the Association of Universities for Research in Astronomy, Inc., under NASA contract NAS 5–26555. These observations are associated with program $\#$16877. 

The data presented herein were obtained at the W. M. Keck Observatory, which is operated as a scientific partnership among the California Institute of Technology, the University of California and the National Aeronautics and Space Administration. The Observatory was made possible by the generous financial support of the W. M. Keck Foundation. The authors wish to recognize and acknowledge the very significant cultural role and reverence that the summit of Maunakea has always had within the indigenous Hawaiian community. We are most fortunate to have the opportunity to conduct observations from this mountain. W. M. Keck Observatory access was supported by Northwestern University and the Center for Interdisciplinary Exploration and Research in Astrophysics (CIERA).

The \textit{HST} data presented in this paper were obtained from the MAST at the Space Telescope Science Institute. The specific observations analyzed can be accessed via \dataset[https://doi.org/10.17909/v0hv-8e68.]{https://doi.org/10.17909/v0hv-8e68}

\vspace{5mm}
\software{\texttt{Astropy} \citep{Astropy13, Astropy18, Astropy22}, \texttt{BLOBCAT} \citep{BLOBCAT}, \texttt{CASA} \citep{casa, CASA22}, \texttt{DrizzlePac} \citep{Avila15}, \texttt{GALFIT} \citep{GALFIT}, \texttt{Matplotlib} \citep{Hunter07}, \texttt{Numpy} \citep{Harris20}, \texttt{PypeIt} \citep{PypeIt}, \texttt{SAOImage DS9} \citep{ds9}, \texttt{Source Extractor} \citep{SExtractor}, \texttt{SciPy} \citep{2020SciPy-NMeth}, \texttt{TinyTim} \citep{Tinytim}}

\facilities{VLA, \textit{HST} (WFC3), Keck (DEIMOS)}
\bibliography{refs}

\end{CJK*}
\end{document}

%% file: aff.tex
\newcommand{\NU}{\affiliation{Center for Interdisciplinary Exploration and Research in Astrophysics (CIERA) and Department of Physics and Astronomy, Northwestern University, Evanston, IL 60208, USA}}

\newcommand{\Swinburne}{\affiliation{Centre for Astrophysics and Supercomputing, Swinburne University of Technology, John St, Hawthorn, VIC 3122, Australia}}

\newcommand{\UCSC}{\affiliation{Department of Astronomy and Astrophysics, University of California, Santa Cruz, CA 95064, USA}}

\newcommand{\STScI}{\affiliation{Space Telescope Science Institute, Baltimore, MD 21218, USA}}

\newcommand{\JHU}{\affiliation{Department of Physics and Astronomy, Johns Hopkins University, Baltimore, MD 21218, USA}}

\newcommand{\PSU}{\affiliation{Department of Astronomy \& Astrophysics, The Pennsylvania State University, University Park, PA 16802, USA}}

\newcommand{\DS}{\affiliation{Institute for Computational \& Data Sciences, The Pennsylvania State University, University Park, PA 16802, USA}}

\newcommand{\GC}{\affiliation{Institute for Gravitation and the Cosmos, The Pennsylvania State University, University Park, PA 16802, USA}}

\newcommand{\CSIRO}{\affiliation{CSIRO, Space and Astronomy, PO Box 76, Epping, NSW 1710, Australia}}

\newcommand{\Fisica}{\affiliation{Instituto de F\'isica, Pontificia Universidad Cat\'olica de Valpara\'iso, Casilla 4059, Valpara\'iso, Chile}}

\newcommand{\Macquarie}{\affiliation{School of Mathematical and Physical Sciences, Macquarie University, NSW 2109, Australia}}

\newcommand{\CfAMacquarie}{\affiliation{Astrophysics and Space Technologies Research Centre, Macquarie University, Sydney, NSW 2109, Australia}}

\newcommand{\Columbia}{\affiliation{Department of Astronomy, Columbia University, New York, NY 10027, USA}}

\newcommand{\THEA}{\affiliation{Theoretical High Energy Astrophysics (THEA) Group, Columbia University, New York, NY 10027, USA}}

\newcommand{\Cahill}{\affiliation{Cahill Center for Astronomy and Astrophysics, California Institute of Technology, Pasadena, CA 91106, USA}}

\newcommand{\DAWN}{\affiliation{Cosmic Dawn Center (DAWN), Denmark}}

\newcommand{\Bohr}{\affiliation{Niels Bohr Institute, University of Copenhagen, Jagtvej 128, DK-2200 Copenhagen N, Denmark}}

\newcommand{\berkeley}{\affiliation{Astronomy Department and Theoretical Astrophysics Center, University of California, Berkeley, Berkeley, CA 94720, USA}}

\newcommand{\NAOJ}{\affiliation{Division of Science, National Astronomical Observatory of Japan,2-21-1 Osawa, Mitaka, Tokyo 181-8588, Japan}}

\newcommand{\IPMU}{\affiliation{Kavli Institute for the Physics and Mathematics of the Universe (Kavli IPMU), 5-1-5 Kashiwanoha, Kashiwa, 277-8583, Japan}}

\newcommand{\ESA}{\affiliation{European Space Agency (ESA), European Space Astronomy Centre (ESAC), Camino Bajo del Castillo s/n, 28692 Villanueva de la Cañada, Madrid, Spain}}

\newcommand{\MIT}{\affiliation{MIT Kavli Institute for Astrophysics and Space Research, Massachusetts Institute of Technology, Cambridge, MA 02139}}

\newcommand{\JIVE}{\affiliation{Joint Institute for VLBI ERIC, Oude Hoogeveensedijk 4, 7991~PD, Dwingeloo, The Netherlands}}

\newcommand{\ASTRON}{\affiliation{ASTRON, Netherlands Institute for Radio Astronomy, Oude Hoogeveensedijk 4, 7991 PD Dwingeloo, The Netherlands}}

\newcommand{\Onsala}{\affiliation{Department of Space, Earth and Environment, Chalmers University of Technology, Onsala Space Observatory, SE-439 92 Onsala, Sweden}}

\newcommand{\McGill}{\affiliation{Department of Physics, McGill University, Montreal, Quebec H3A 2T8, Canada}}

\newcommand{\Curtin}{\affiliation{International Centre for Radio Astronomy Research, Curtin University, Bentley, WA 6102, Australia}}

\newcommand{\UAmsterdam}{\affiliation{Anton Pannekoek Institute for Astronomy, University of Amsterdam, Science Park 904, 1098 XH, Amsterdam, The Netherlands}}

%% file: authors.tex
\author[0000-0002-9363-8606]{Yuxin Dong (董雨欣)}
\NU

\author[0000-0003-0307-9984]{Tarraneh Eftekhari}
\altaffiliation{NHFP Einstein Fellow}
\NU

\author[0000-0002-7374-935X]{Wen-fai Fong}
\NU

\author[0000-0001-9434-3837]{Adam T. Deller}
\Swinburne

\author{Alexandra G. Mannings}
\UCSC

\author[0000-0003-3801-1496]{Sunil Simha}
\UCSC

\author[0000-0002-5519-9550]{Navin Sridhar}
\Columbia
\THEA
\Cahill

\author[0000-0002-9946-4731]{Marc Rafelski}
\STScI
\JHU

\author[0000-0002-5025-4645]{Alexa C. Gordon}
\NU

\author[0000-0003-3460-506X]{Shivani Bhandari}
\CSIRO

\author[0000-0002-8101-3027]{Cherie K. Day}
\McGill

\author[0000-0002-9389-7413]{Kasper E. Heintz}
\DAWN
\Bohr

\author[0000-0003-2317-1446]{Jason W.T. Hessels}
\ASTRON
\UAmsterdam

\author[0000-0001-6755-1315]{Joel Leja}
\PSU
\DS
\GC

\author[0000-0002-6437-6176]{Clancy W. James}
\Curtin

\author[0000-0002-5740-7747]{Charles~D.~Kilpatrick}
\NU


\author[0000-0002-5053-2828]{Elizabeth K. Mahony}
\CSIRO

\author[0000-0001-9814-2354]{Benito~Marcote} 
\JIVE

\author[0000-0001-8405-2649]{Ben Margalit}
\berkeley

\author[0000-0003-0510-0740]{Kenzie Nimmo}
\MIT

\author[0000-0002-7738-6875]{J. Xavier Prochaska}
\UCSC
\IPMU
\NAOJ

\author[0000-0003-3937-0618]{Alicia Rouco Escorial}
\altaffiliation{ESA Research Fellow}
\ESA

\author[0000-0003-4501-8100]{Stuart D. Ryder}
\Macquarie
\CfAMacquarie

\author[0000-0001-9915-8147]{Genevieve Schroeder}
\NU

\author[0000-0002-7285-6348]{Ryan~M.~Shannon}
\Swinburne

\author[0000-0002-1883-4252]{Nicolas Tejos}
\Fisica

%% file: tables/radio_obs.tex
\begin{deluxetable*}{ccccccc}[t!]
\linespread{1.2}
\tablecaption{Imaging Results from VLA Observations of the FRB20201124A Source}
\tablecolumns{7}
\tablewidth{0pt}
\label{tab:radio_obs}
\tablehead{
\colhead{Frequency} &
\colhead{Bandwidth} & 
\colhead{Pixel Scale} &
\colhead{Image RMS} &
\colhead{Beam Size} &
\colhead{Beam Angle} &
\colhead{Flux Density} \\
\colhead{(GHz)} &
\colhead{(GHz)} &
\colhead{$\arcsec$/pix} &
\colhead{($\mu$Jy)} &
\colhead{(arcsec$^2$)} &
\colhead{(deg)} &
\colhead{($\mu$Jy)}
}
\startdata
1.5 & 0.8 & 0.26 & 13.95 & 1.40 $\times$ 1.31 & -13.39 & $750 \pm 114$ \\
3 & 1.6 & 0.13 & 6.29 & 0.68 $\times$ 0.65 & -14.51 & $547 \pm 82$ \\
6 & 3.6 & 0.066 & 1.94 & 0.37 $\times$ 0.36 & -58.23 & $366 \pm 55$\\
\enddata
\end{deluxetable*}

%% file: tables/sfr.tex
\begin{deluxetable}{ccc}[t!]
\linespread{1.2}
\tablecaption{Global Host Star Formation Rates for FRB\,20201124A \label{tab:sfr}}
\tablecolumns{3}
\tablewidth{0pt}
\tablehead{
\colhead{Method} &
\colhead{SFR} &
\colhead{Reference} \\
\colhead{} &
\colhead{(M$_{\odot}$ yr$^{-1}$)}
}
\startdata
\multicolumn{3}{c}{\emph{{\tt Prospector} Spectral Energy Distribution Fitting}} \\
\hline
Parametric & $2.43 \pm 0.13$ & \cite{Fong21} \\
Parametric & $4.3 \pm 0.5$ & \cite{Piro21} \\
Non-Parametric & $1.50^{+0.52}_{-0.35}$ & \cite{Fong21} \\
Non-Parametric & $2.72^{+1.65}_{-1.22}$ & \cite{Gordon2023} \\
\hline
\multicolumn{3}{c}{\emph{Luminosity-SFR relationship}} \\
\hline
Optical (H$_{\alpha}$) & $2.1^{+0.7}_{-0.3}$ & \cite{Fong21} \\
Optical (H$_{\alpha}$) & $1.7 \pm 0.6$ & \cite{Ravi22} \\
Optical (H$_{\alpha}$) & $3.4 \pm 0.3$ & \cite{Xu22} \\
IR (3-1100 $\mu$m) & $4.0^{+0.9}_{-0.5}$ & \cite{Fong21} \\
Radio (1.4 GHz) & 2.2 -- 5.9 & \cite{Fong21} \\
Radio (1.4 GHz) & 10 & \cite{Piro21} \\
Radio (1.4 GHz) & 7 & \cite{Ravi22} \\
Radio (1.5 GHz) & $8.9 \pm 0.7$  & This Work 
\enddata
\end{deluxetable}

%% file: tables/radio_opt_SFR.tex
\setlength{\tabcolsep}{6pt} 
\renewcommand{\arraystretch}{1.5} 
\begin{deluxetable*}{l|ccccc}[!t]
\tabletypesize{\scriptsize}
\tablecaption{Spectral Luminosities, Radio and Optical SFRs of FRBs from the Literature}
\tablecolumns{5}
\tablewidth{0pc}
\centering
\tablehead{
\colhead {FRB}	 &
\colhead{$\nu$} &
\colhead {L$_\nu$}  &
\colhead {SFR$_{\rm radio}$} &
\colhead{SFR$_{\rm opt}$} \\
\colhead{} &
\colhead{(GHz)} &
\colhead{(ergs$^{-1}$Hz$^{-1}$)} &
\colhead{(M$_{\odot}$ yr$^{-1}$)} &
\colhead{(M$_{\odot}$ yr$^{-1}$)}
}

\label{tab:all_sfr}
\startdata
171020A & 2.1 & $<$3.2$\times 10^{28}$ & $<$2.65 & 0.13$^a$ \\
180301A & 1.5 & $<$1.8$\times 10^{29}$ & $<$10.82 & 1.93$^b$ \\
180916B & 1.7 & $<$4.9$\times 10^{26}$ & $<$0.03 & 0.06 \\
180924A & 6.5 & $<$5.7$\times 10^{28}$ & $<$9.58 & 0.88 \\
181030A & 3.0 & 1.3$\times 10^{27~c}$ & 0.08 & 0.35$^c$ \\
181112A & 6.5 & $<$1.3$\times 10^{29}$ & $<$21.14 & 0.37 \\
190102A & 6.5 & $<$4.2$\times 10^{28}$ & $<$7.11 & 0.86 \\
190523A & 3.0 & $<$4.3$\times 10^{30}$ & $<$392.89 & $<$0.09 \\
190611A & 0.9 & $<$2.9$\times 10^{30}$ & $<$120.63 & 0.69 \\
190608B & 5.5 & 2.5$\times 10^{28~d}$ & 3.95 & 7.03$^{e}$ \\
190711A & 0.9 & $<$5.6$\times 10^{30}$ & $<$226.09 & 0.42 \\
190714A & 3.0 & $<$7.2$\times 10^{29}$ & $<$71.86 & 0.65 \\
191001A & 2.0 & 4.2$\times 10^{29~f}$ & 34.60 & 18.28$^{e}$ \\
191228A & 6.5 & $<$3.4$\times 10^{28}$ & $<$5.82 & 0.50$^b$ \\
200120E & 8.4 & $<$3.8$\times 10^{25~g}$ & $<$0.008 & 0.6$^{h}$\\
200428A & $\cdots$ & $\cdots$ & 2$^i$ & 1.65$^{h}$ \\
200430A & 3.0 & $<$3.2$\times 10^{29}$ & $<$32.56 & 0.26$^b$ \\ 
200906A & 6.0 & $<$4.3$\times 10^{28}$ & $<$6.76 & 0.48$^b$ \\
210117A & 5.0 & $<$5.3$\times 10^{28~j}$ & $<$7.61 & 0.01$^{j}$ \\
220207C & 3.0 & $<$2.3$\times 10^{28}$ & $<$2.42 & 2.14$^k$ \\
220307B & 3.0 & $<$1.0$\times 10^{30}$ & $<$99.53 & 3.52$^k$ \\
220310F & 3.0 & $<$4.6$\times 10^{30}$ & $<$435.21 & 0.15$^k$ \\
220319D & 1.4 & $<$1.8$\times 10^{26}$ & $<$0.01 & 1.17$^k$ \\
220418A & 3.0 & $<$8.6$\times 10^{30}$ & $<$791.26 & 0.37$^k$ \\
220506D & 3.0 & $<$1.5$\times 10^{30}$ & $<$147.47 & 7.01$^k$ \\
220509G & 3.0 & $<$1.1$\times 10^{29}$ & $<$11.40 & 0.08$^k$ \\
220825A & 3.0 & $<$9.4$\times 10^{29}$ & $<$93.71 & 1.34$^k$ \\
220914A & 3.0 & $<$1.8$\times 10^{29}$ & $<$18.54 & 1.45$^k$ \\
220920A & 3.0 & $<$3.6$\times 10^{29}$ & $<$36.64 & 0.39$^k$ \\
221012A & 3.0 & $<$1.4$\times 10^{29}$ & $<$138.14 & 0.49$^k$ \\
\hline
121102A & 1.7 & $<$2$\times 10^{29~l}$ & $<$13.55 & 0.4$^m$ \\
190520B & 1.7 & $<$3$\times 10^{29~n}$ & $<$34.60 & 0.04$^e$ \\
\enddata
\tablecomments{Unless otherwise specified, all luminosity measurements are from \cite{Law22, Law23} and all optical SFRs are from \cite{Heintz20}. We add the the luminosities of PRS\,20121102A and 20190520B as upper limits on star formation from \cite{Marcote17} and \cite{Bhandari23b}, respectively. While FRB\,20190714A has detectable radio emission, it is not clear at present whether it is due to star formation or is a compact PRS \citep{Chibueze22}. Thus, we keep the upper limit from \cite{Law22}.\\}
$^a$ \footnotesize{\citet{Mahony18}}
$^b$ \footnotesize{\citet{Bhandari22c}}
$^c$ \footnotesize{\citet{Bhardwaj21}}
$^d$ \footnotesize{\citet{Bhandari20a}}
$^e$ \footnotesize{\cite{Gordon2023},}
$^f$ \footnotesize{\citet{Bhandari20b},}
$^g$ \footnotesize{\citet{Miller10},}
$^h$ \footnotesize{\citet{Hatsukade22},} 
$^i$ \footnotesize{\citet{Chomiuk11},} 
$^j$ \footnotesize{\citet{Bhandari23}}
$^k$ \footnotesize{\citet{Law23}}
$^l$ \footnotesize{\citet{Marcote17}}
$^m$ \footnotesize{\citet{Tendulkar2017}}
$^n$ \footnotesize{\citet{Bhandari23b}}
\end{deluxetable*}